\documentclass[useAMS,usenatbib]{mn2e}
\usepackage[draft]{hyperref} 
\usepackage{ulem}
\usepackage{color}
\usepackage{afterpage}
\usepackage[utf8]{inputenc}
\usepackage{graphicx,url,times}
\usepackage{multirow} 
\usepackage{booktabs,threeparttable}
\newif\ifpdf
  \ifx\pdfoutput\undefined
    \pdffalse
  \else
    \ifnum\pdfoutput=1
      \pdftrue
    \else
      \pdffalse
    \fi
  \fi

\ifpdf
    \graphicspath{{plots/}}
\else
    \graphicspath{{eps/}}
\fi

\newcommand{\papertitle}{Sussing Merger Trees: Stability and Convergence}

\ifpdf\hypersetup{
pdftitle={\papertitle},
pdfauthor={Yang Ocean Wang},
pdfkeywords={N-body simulations, haloes evolution, dark matter},
pdfstartview=FitH,
}
\fi

\newcommand{\consistenttree}{\textsc{Consistent Trees}}
\newcommand{\dtree}{\textsc{D-Trees}}
\newcommand{\mergertree}{\textsc{MergerTree}}

\newcommand{\hbt}{\textsc{HBT}}
\newcommand{\jmerger}{\textsc{JMerge}}
\newcommand{\subfind}{\textsc{Subfind}}
\newcommand{\sublink}{\textsc{SubLink}}
\newcommand{\treemaker}{\textsc{TreeMaker}}
\newcommand{\velociraptor}{\textsc{VELOCIraptor}}
\newcommand{\ysamtm}{\textsc{ySAMtm}}
\newcommand{\TO}{\textsc{TO}}
\newcommand{\MO}{\textsc{MO}}
\newcommand{\EO}{\textsc{EO}}
\newcommand{\QO}{\textsc{QO}}
\newcommand{\LO}{\textsc{LO}}
\newcommand{\class}{\textsc{Class2}}

\newcommand{\hires}{\textsc{HiRes}}
\newcommand{\lowres}{\textsc{LoRes}}
\newcommand{\singlehalo}{\textsc{SingleHalo}}
\newcommand{\halos}{haloes}
\newcommand{\Halos}{Haloes}
\newcommand{\hkpc}{{\ifmmode{h^{-1}{\rm kpc}}\else{$h^{-1}$kpc}\fi}}

\newcommand{\Fig}[1]{Figure~\ref{#1}}

\newlength{\figwidth}
\newlength{\resplot}
\begin{document}
\title[Stability and Convergence of Merger Trees]{\papertitle}
\author[Yang Wang et al.]{Yang
	Wang,$^{1,2,3,4}$\thanks{wangyang23@mail.sysu.edu.cn}
  Frazer R. Pearce,$^{2}$
  Alexander Knebe,$^{5,6}$
  Aurel Schneider,$^{7,8}$
  \newauthor
  Chaichalit Srisawat,$^{7}$
  Dylan Tweed,$^{9,10}$
  Intae Jung,$^{11}$
  Jiaxin Han,$^{12}$
  \newauthor
  John Helly,$^{12}$
  Julian Onions,$^{2}$
  Pascal J. Elahi,$^{13}$
  Peter A. Thomas,$^{7}$
  \newauthor
  Peter Behroozi,$^{14,15}$
  Sukyoung K. Yi,$^{16}$
  Vicente Rodriguez-Gomez,$^{17}$
  \newauthor
  Yao-Yuan Mao,$^{14,15}$
  Yipeng Jing,$^{10}$
  and Weipeng Lin$^{1,3}$\\
$^{1}${School of Physics and Astronomy, Sun Yat-Sen University, Guangzhou, 510275, China}\\
$^{2}${School of Physics \& Astronomy, University of Nottingham, Nottingham, NG7 2RD, UK}\\
$^{3}${Key Laboratory for Research in Galaxies and Cosmology, Shanghai Astronomical Observatory, Shanghai 200030, China}\\
$^{4}${Graduate School of the Chinese Academy of Science, 19A, Yuquan Road, Beijing, China}\\
$^{5}${Departamento de F\'isica Te\'{o}rica, M\'{o}dulo 15, Facultad de Ciencias, Universidad Aut\'{o}noma de Madrid, 28049 Madrid, Spain}\\
$^{6}${Astro-UAM, UAM, Unidad Asociada CSIC}\\
$^{7}${Department of Physics and Astronomy, University of Sussex, Brighton BN1 9QH, UK}\\
$^{8}${Institute for Computational Sciences,University of Zurich, Switzerland}\\
$^{9}${Racah Institute of Physics, The Hebrew University, Jerusalem 91904, Israel}\\
$^{10}${Center for Astronomy and Astrophysics, Department of Physics, Shanghai Jiao Tong University, Shanghai 200240, China}\\
$^{11}${Department of Astronomy, The University of Texas at Austin, Austin, TX, 78712, USA}\\
$^{12}${Institute for Computational Cosmology, Department of Physics, Durham University, South Road, Durham DH1 3LE, UK}\\
$^{13}${Sydney Institute for Astronomy, A28, School of Physics, The University of Sydney, NSW 2006, Australia}\\
$^{14}${Kavli Institute for Particle Astrophysics and Cosmology and Physics Department, Stanford University, Stanford, CA 94305, U SA}\\
$^{15}${SLAC National Accelerator Laboratory, Menlo Park, CA 94025, USA}\\
$^{16}${Department of Astronomy and Yonsei University Observatory, Yonsei University, Seoul 120-749, Republic of Korea}\\
$^{17}${Harvard-Smithsonian Center for Astrophysics, 60 Garden Street, Cambridge MA, 02138, USA}\\
}

\date{Accepted XXX. Received XXX; in original form XXX}

\pagerange{\pageref{firstpage}--\pageref{lastpage}} \pubyear{2013}\volume{0000}

\maketitle
\label{firstpage}

\begin{abstract}
Merger trees are routinely used to follow the growth and merging history of
dark matter \halos\ and sub\halos\ in simulations of cosmic structure
formation.  \citet{Srisawat13} compared a wide range of
merger-tree-building codes. Here we test the influence of output
strategies and mass resolution on tree-building. We find that, somewhat
surprisingly, building the tree from more snapshots does not generally
produce more complete trees; instead, it tends to shorten
them. Significant improvements are seen for patching schemes which
attempt to bridge over occasional dropouts in the underlying halo
catalogues or schemes which combine the halo-finding and tree-building
steps seamlessly.  The adopted output strategy does not affect the
average number of branches (bushiness) of the resultant merger trees. However, mass
resolution has an influence on both main branch length and the
bushiness. As the resolution increases, a halo with the same mass can
be traced back further in time and will encounter more small
progenitors during its evolutionary history.  Given these results, we
recommend that, for simulations intended as precursors for galaxy
formation models where of order 100 or more snapshots are analysed,
the tree-building routine should be integrated with the halo finder,
or at the very least be able to patch over multiple adjacent
snapshots.
\end{abstract}


\section{Introduction} \label{sec:introduction}
In the current standard cosmological model, galaxies are thought to
form and evolve within the potential well of a surrounding dark matter
halo \citep{white_1978,E&S_1983,Blum_1984}. Gas assembly and star
formation takes place within this environment, and the hierarchical
merging of the \halos\ gives rise to galaxy mergers. The whole life of
a galaxy is intimately connected to this underlying host halo
framework.

Thus, when undertaking galaxy modeling, a set of appropriate \halos\
and their associated merger tree histories are a key ingredient. One leading
approach for this task is using Semi-Analytic Models (SAMs). The
merger trees employed by SAMs can be derived from the Press-Schechter
formalism \citep{PS_1974} or extended Press-Schechter formalism
\citep{Bond_1991}, or in a more realistic way directly from $N$-body
simulations (see \cite{1997MNRAS.292..835R, Lacey_1993} for the
historical origin of both approaches). The latter approach has become
popular since $N$-body simulations can provide more realistic halo
histories in complex environments. Although merger trees derived
from $N$-body simulations are widely used, their performance and
properties have not been thoroughly studied to date. This was the aim
of our \textsc{Sussing Merger Trees}
workshop\footnote{http://popia.ft.uam.es/SussingMergerTrees} from
which arose a series of comparisons studying various aspects of merger
trees.

\cite{Srisawat13} was the first paper from the \textsc{Sussing Merger
  Trees} workshop. It gave a general overview on the contributing
merger-tree-building methods. As we found, different tree codes
produce distinctly different results. Following up this work,
\cite{Lee_2014} found that for SAMs, the $z=0$ galaxy properties are
altered if different halo merger-tree-building algorithms are
used. The star formation history and the properties of satellite
galaxies can be remarkably different. They also showed that these
changes could be largely alleviated if the model was re-tuned to the
input tree.  This work demonstrated that different tree
building algorithms construct different merger trees, which cannot all
reflect the true structure of the underlying dark matter halo
framework they are purporting to encapsulate. Thus although re-tuning
is possible this is not an ideal situation. In this paper we attempt
to quantify the differences between the different algorithms by
varying both the simulation output strategy and the simulation
resolution.

Besides this general view, we are also interested in any aspects that
will affect merger tree construction. Basically there are two steps
for building merger trees. First we need an input halo catalogue which
is usually found by employing a halo finding code on every
snapshot. Then these catalogues of \halos\ from each snapshot are
linked together by a tree-building code to construct a merger
tree. Since the input halo catalogue can be varied when different halo
finding codes are applied, it is natural that the input halo catalogue
can affect the final merger trees. \cite{Avila_2014} found that the
underlying halo finder is very relevant to the merger trees
built. Different underlying halo catalogues result in changes to the
main branch length and the branching ratio of the resultant merger
trees.

In this work, we explore fundamental aspects that influence merger
tree construction. We test whether our 9 contributing tree-building
algorithms can recover stable and convergent merger trees when the
simulation output strategy is changed and discuss an optimal strategy.
This work contains two parts: first we test merger tree stability by
changing the output frequency of the underlying simulation; we follow
this by testing convergence by changing the mass resolution of the
simulation. We study the performance of the various merger tree
builders under these changes in the numerical input conditions.

The rest of this paper is organized as follows: we start by describing
the simulation data we have used and list the merger tree building
codes applied in Section 2. We then define the properties we measure
for our merger trees in Section 3. Our main results are given in
Sections 4, 5 and 6, followed by a discussion and some conclusions in
Section 7.


\begin{figure*}
    \centering
    \includegraphics[angle=-90,width=\linewidth,totalheight=80mm]{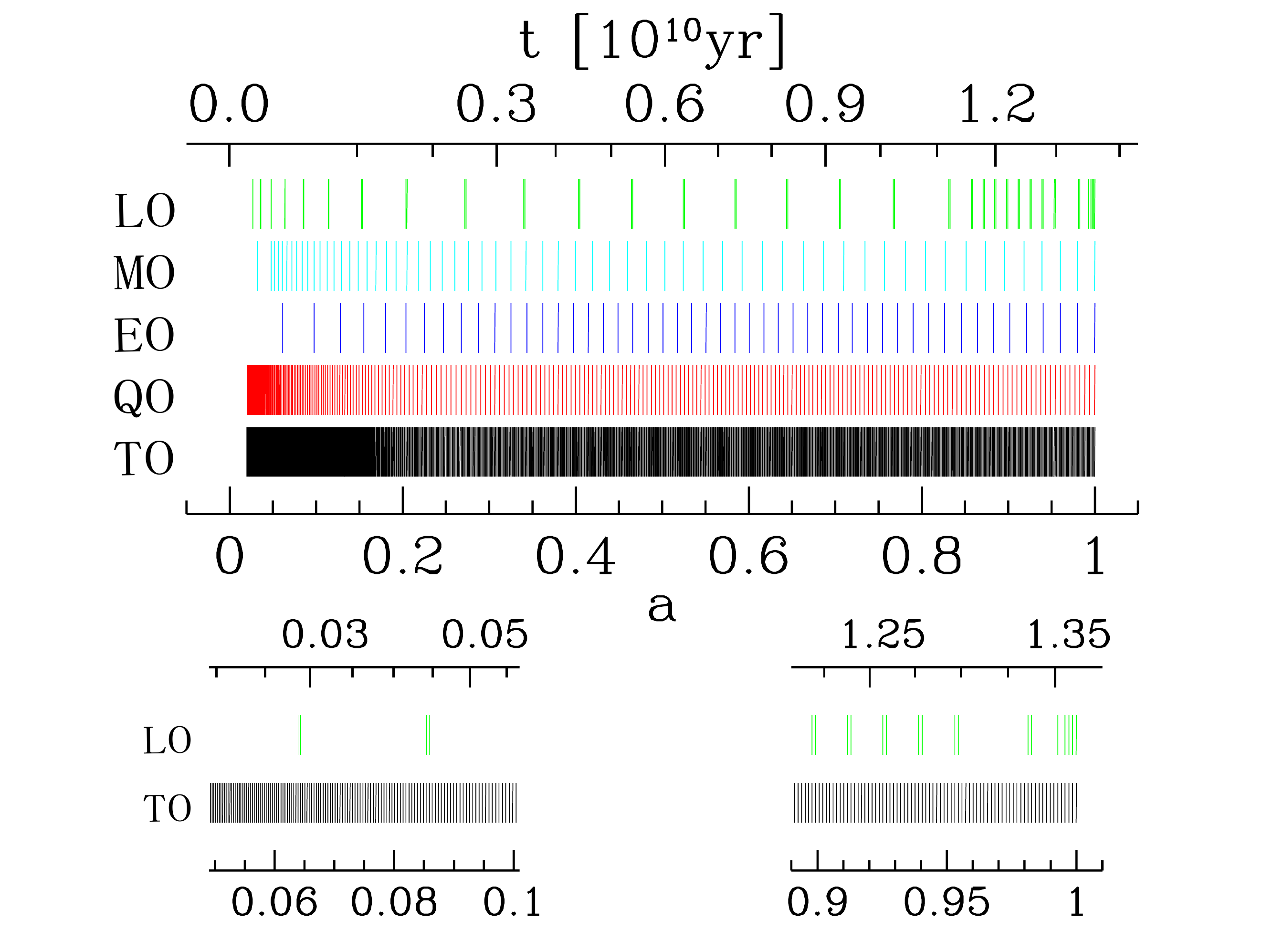}
    \caption{Output scheme for all five datasets. The horizontal axis
      lists both the expansion factor $a$ and cosmic time $t$. A
      vertical bar in a specific position indicates a snapshot at the
      corresponding time. Each set of vertical lines illustrates one
      dataset, as listed on the left. In the lower part, two zoom in
      regions are shown for the Thousand Output(\TO) and Lame
      Output(\LO) datasets.}
    \label{pic:output}
\end{figure*}

\section{Simulation Data and Merger Tree Builders} \label{sec:sim}
\subsection{Stability study}
The first simulation is a dark-matter-only re-simulation of a Milky
Way-like halo taken from the Aquarius project
\citep[hereafter \singlehalo]{springel_aquarius_2008}. Specifically we
use Aquarius halo A at level 4, which has a mass resolution of
$M_{\rm{p}}=2.868\times10^5 {\rm M_\odot}$. For the Aquarius project the
underlying cosmology is $\Omega_M=0.25$, $\Omega_{\Lambda}=0.75$,
$\sigma_8=0.9$, $h=0.73$. Because this simulation is a single halo re-simulation, most of
the \halos\ at redshift 0 (which will be the root \halos\ of our merger trees)
 are sub\halos. Only a few of them are distinct \halos.

In total the simulation outputs 1024 snapshots (given IDs 0 to 1023
sequentially from the earliest snapshot to the final one) from
redshift 50 to 0. We subsample the full output set in order to
undertake our merger tree stability study. This study attempts to
quantify the effects of varying the output strategy on the resulting
merger tree. To test this, we extract specific snapshots from the
original output set of the \singlehalo\ simulation. There are 5 sets
for analysis:
\begin{itemize}
\item Using all 1024 snapshots. This forms the Thousand Output (\TO)
dataset.

At the beginning of the simulation the time interval between output
snapshots is about $4.5\times10^5$ years. This slowly increases with
time and reaches $1.9\times10^7$ years at snapshot 440 (where the
redshift of 2.93). After that the time interval remains constant at
$1.9\times10^7$ years.

\item Using every 4th snapshot from the full \TO\ set, i.e. the 4th, 8th,
12th ... 1024th snapshot. There are 256 snapshots in this set, which
forms the Quarter Output (\QO) dataset.

\item Selecting 64 outputs matching the Millennium simulation output
  strategy, which are equally spaced in log expansion factor at high
  redshift. This set forms the Millennium Output (\MO) dataset. The
  time interval for this set is $\Delta \ln a \approx 0.081$, where
  $a$ is the expansion factor, at high redshift.  It gradually
  decreases with time and reaches a value of $\Delta \ln a \approx
  0.020$ by $z=0$.

\item Selecting 64 outputs equally spaced in time. This set forms the
Equally timed Output (\EO) dataset. The time interval for this set
is about $2.7\times10^8$ years.

\item A set of 64 outputs deliberately selected to be a poor choice. In
this case we select pairs of adjacent snapshots followed by a large
gap, particularly at early times. At late times the gaps reduce and
there are more neighbouring snapshots. This set forms the Lame Output
(\LO) dataset.
\end{itemize}
These five output strategies are displayed visually in
\Fig{pic:output}, which illustrates the spacing of the various
snapshots as a function of cosmic time and expansion factor. Expanded
timelines for the \TO\ and \LO\ datasets are inserted to illustrate the
many snapshots in the base \TO\ simulation and the pairing of snapshots
in the \LO\ dataset.

\subsection{Convergence study}
To study the behaviour of merger trees with mass resolution we have
also run two dark matter-only simulations with the same initial
conditions and cosmology, but at a different resolution. They each follow the
evolution of structure in a small box of comoving side $8 h^{-1}$ Mpc
containing $512^3$ particles (hereafter the \hires\ simulation) and
$256^3$ particles (hereafter the \lowres\ simulation),
respectively. The resolution is $m_p = 2.1\times 10^6\,h^{-1}{\rm
M_\odot}$ for \lowres\ simulation and $m_p = 2.6\times 10^5\,h^{-1}{\rm
M_\odot}$ for \hires\ simulation.
Their mass resolution is equivalent to the resolution of
the Aquarius project simulations at level 5 and level 4, respectively
\citep{springel_aquarius_2008}. The mass resolution of the Aquarius simulation
at level 5 is roughly three times higher
than that of the Millennium-II simulation ($m_p = 6.9\times 10^6\,h^{-1}{\rm
M_\odot}$) used by \citet{lee_2013}. The cosmology was chosen to be
the same as in the Aquarius simulation, i.e. $\Lambda$CDM with
$\Omega_M = 0.25,\: \Omega_\Lambda = 0.75,\: \sigma_8 =0.9,\: n_s =
1,\: h = 0.73$.  Initial conditions were generated at $z=127$ using
the Zel'dovich approximation to linearly evolve positions from an
initially glass-like state. This was then evolved to the present day
using \textsc{gadget}-2 \citep{springel_cosmological_2005} with a
gravitational softening equal to $0.04\hkpc$. Both simulations output 51
snapshots. We study the appearance of the merger trees that result
from these two simulations which only differ in mass resolution.

In contrast to the stability study, the two simulations used in
our convergence study are full box cosmological simulations. They contain
populations of both main \halos\ and sub\halos. Since the convergence
of merger trees concerns mergers of \halos, cosmological
simulations are more appropriate for this study.
\begin{figure*}
    \centering
    \includegraphics[width=1\linewidth]{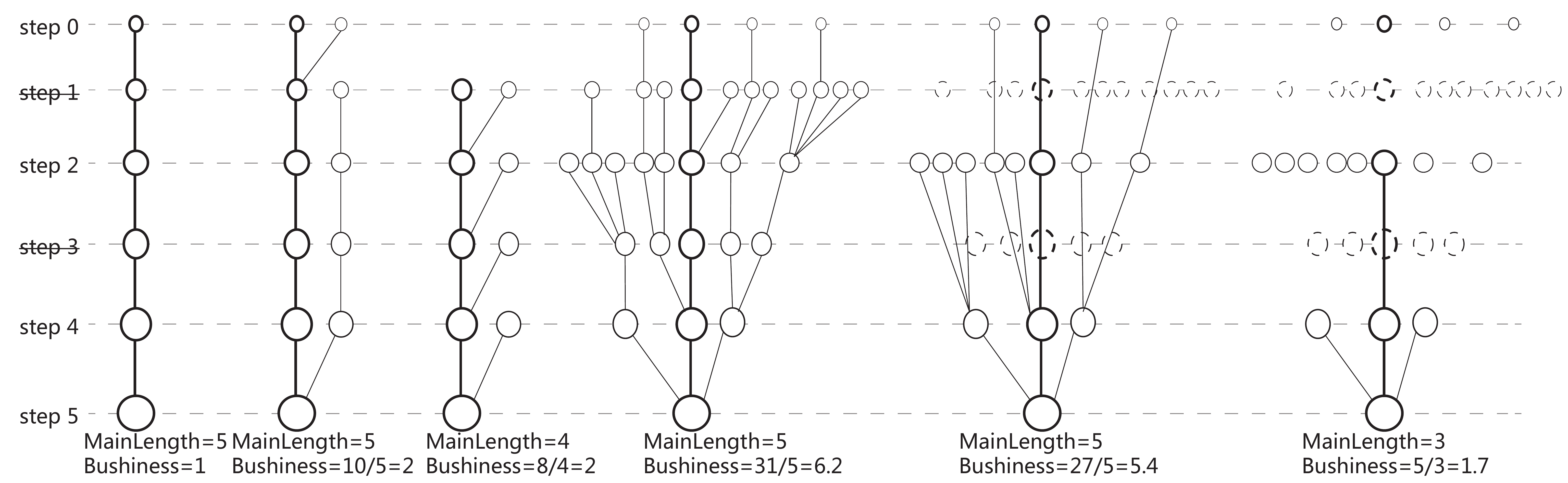}
    \caption{Calculation of main branch length {\it L} and bushiness
      {\it B} as defined in the text for 6 example trees. The thick
	  line represents the main branch. If snapshots
      are dropped to form a smaller dataset then the fourth tree can
      become the fifth or the sixth depending on whether or not the
      progenitor branches are well linked. The sixth tree has a
      significantly reduced bushiness.}
    \label{pic:illus}
\end{figure*}
\subsection{Description of tree builders} \label{sec:treebuilder}

We employ nine different algorithms to build merger trees in this
work. In this section we briefly list the contributed tree
builders. For this work we supplied a single (sub)halo catalogue
generated by \subfind\ \citep{subfind_2001} in order to ensure
consistency for the halo generation step. A comparison of the effect
of different halo finding algorithms on the resultant trees has
already been completed by \cite{Avila_2014}. We asked each of the tree
building teams to construct merger trees based on this input halo
catalogue. \hbt\ and to some extent \consistenttree\ are slightly
different from other tree-builders on this point in that they add
sub\halos\ while building the merger tree, and in effect edit the
halo population.

All of the tree-builders except \jmerger\ trace the \halos\ via individual
particle IDs which are matched between snapshots. They link two \halos\
together if they share the same particles and can meet the requirement
of a merit function. Their merit functions are somewhat different
(refer to \cite{Srisawat13} for further details). The nine supplied
algorithms split into 4 broad types:
\begin{itemize}
\item {\bf Class 1:} Example: \jmerger\ (Onions). This simplest type aims to
build merger trees from simulation datasets which do not include
particle IDs. They only require the halo mass and trajectory to match
\halos\ with their progenitors.

\item {\bf Class 2:} Examples: \mergertree\
\citep{knollmann_ahf:_2009, gill_2004b}, \treemaker\ \citep{2009A&A...506..647T},
\velociraptor\ \citep{2011MNRAS.418..320E}, \ysamtm\ \citep{2014ApJ...794...74J}. The simplest tree-building
algorithms that make use of particle IDs to assist halo matching between
snapshots. They only use adjacent snapshots and do not attempt to
correct the resultant trees for any defects in the halo catalogue or
halo dropouts. Since their results are almost indistinguishable, we label them as \class\ codes.

\item {\bf Class 3:} Examples: \sublink\ \citep{2015MNRAS.449...49R}, \dtree\
\citep{2014MNRAS.440.2115J}, \consistenttree\ \citep{2013ApJ...762..109B}.  These are more
sophisticated algorithms that attempt to patch the constructed trees by
searching for matches over several snapshots. They will search one or
more steps further if they can't find the progenitors in consecutive
snapshots. \consistenttree\ will also insert `fake' progenitors when
it feels that there are missing ones, and it uses additional
information such as the gravitational motion of \halos\ in determining
progenitor matches.

\item {\bf Class 4:} Example: \hbt\ \citep{HBT_2012}. This class of
  tree builder tracks \halos\ and sub\halos\ throughout the
  simulation, intimately connecting the halo finding step with the
  tree-building stage. First it takes a friends-of-friends halo
  catalogue as an input catalogue.  It tracks \halos\ from the first
  snapshot in which they appear, and adds their descendants to the
  subhalo population if they are accreted and survive.
\end{itemize}
For a detailed description of each of these tree-building codes please
refer to section 4 of \cite{Srisawat13} and the appropriate individual
methods papers.


\section{Analysed merger tree properties}
\subsection{Merger tree geometry}

Quantifying the merger tree geometry is fundamental to further
analysis. Such physical properties are derived from the structure of
the merger tree.  In this work, two main properties are used to
describe the merger tree geometry: the length of the main branch
(hereafter {\it L}) and the average number of branches of the merger tree
(hereafter referred to as the bushiness, {\it B}). In previous work in
this series, the main branch length and the branching ratio (the
number of direct progenitors of a halo) were used to characterize a
merger tree. Here the branching ratio is replaced with bushiness,
because the former will introduce bias to any comparison between
merger trees with different numbers of snapshots and snapshot
intervals. A uniform parameter, such as bushiness, circumvents this
issue.

The length of the main branch illustrates how far a halo can be traced
back in time through a succession of snapshots. In this work, {\it L}
is defined as the difference between the IDs of the snapshot of the
root and the snapshot of the earliest main progenitor. Note that we
fix the snapshot IDs within the full \TO\ dataset and preserve these
numbers when selecting a subset of snapshots to produce the other four
datasets that make up the stability analysis. A detailed definition of
what forms the main branch and other terminology used here can be
found in section 2 of \cite{Srisawat13}.

The bushiness of a merger tree is a measure of its average number of branches. It
is defined as the sum of the length of all the connections that form
the tree divided by the length of the main branch. A connection is a
link between two nodes, i.e. a link between a halo in the merger tree
and one of its direct progenitors. The difference between the IDs of
the snapshots which contain the two nodes is the length of this
connection. As before, the original snapshot ID from the \TO\ dataset
is used in this calculation so that this parameter can be compared
among different sets.

\Fig{pic:illus} illustrates the calculation of main branch length and
bushiness. The fifth and sixth trees (from the left) show how the main
branch length and bushiness of the fourth tree changes when the output
strategy is changed by dropping the indicated snapshots. Assuming that
snapshots 1 and 3 are dropped from the tree, if a tree builder
retained the ability to link progenitors despite the missing snapshot,
the fourth tree will become the fifth. However, if progenitor branches
cannot be accurately linked the tree can collapse to the sixth tree.
The main branch is shown as the thick line in the middle of
each tree.

\Fig{pic:illus} illustrates how the average tree bushiness is
calculated in practice. For example, the 4th tree has 31 connections
all of length unity and a main branch length of 5, resulting in a
bushiness of $31/5=6.2$. For the fifth tree, the tree builder has
worked well and reconstructed a structure similar to the full tree.
There are 15 connections, 12 of length two and 3 of unit length, for a
total value of 27. The main branch retains length 5 and the bushiness
is almost unchanged at 5.4. However, for the much reduced sixth tree,
there are only 4 recovered connections, one of length two and three of
unit length, for a total value of 5. The main branch length is also
smaller at 3. The bushiness of this tree is hence reduced to 1.7. In
general, a merger tree with larger bushiness has more connections with
respect to its main branch and appears wider and more complex. A low
bushiness implies a thin tree, dominated by the main branch.

\subsection{Physical properties}

The mass assembly history (MAH) of \halos\ provides an important
constraint on models of galaxy formation. We investigate the build-up
of mass within the trees as a function of cosmic time by measuring the
total mass within tree main branches at each snapshot, normalized to
the mass at $z=0$. This parameter illustrates how far back the mass
assembly history can be
traced \cite[see also,][]{2014MNRAS.440.2115J}.

The merger rate is another important physical property in galaxy
formation. A galaxy's shape, metallicity and colour may be affected by
merger events. Thus, as they form the underlying framework, the
mergers of \halos\ also need to be considered. In this work, we study
both mergers between \halos\ and mergers between
sub\halos. \Halos\ more massive than $10^8{\rm M_{\odot}}$ are taken
into account. We calculate the mean merger number per descendant halo
per $\rm Gyr$ as a function of redshift. This parameter has been
widely used in many works investigating the mergers of dark matter
\halos\ and galaxies \cite[see, e.g.,][]{Fakhouri2008, Fakhouri2009,
  Fakhouri2010, Genel2009, Genel2010}.

It should be clarified that, particularly when aggregated, the merger
rate and the bushiness are related but not identical properties of a set
of merger trees. A larger bushiness implies that a merger tree has more
sub-branches per unit length, while a larger merger rate similarly
implies more mergers per unit time. However, the precise time-step at
which a merger is counted varies between the two properties because
subhalo trees are not connected to the host halo tree until the
subhalo has been effectively destroyed by the host halo. Hence the
number of progenitors is not directly equivalent to the number of
mergers on any particular step.


\section{Result I: Consistency Test}
\begin{figure}
    \centering
    \includegraphics[width=0.85\linewidth]{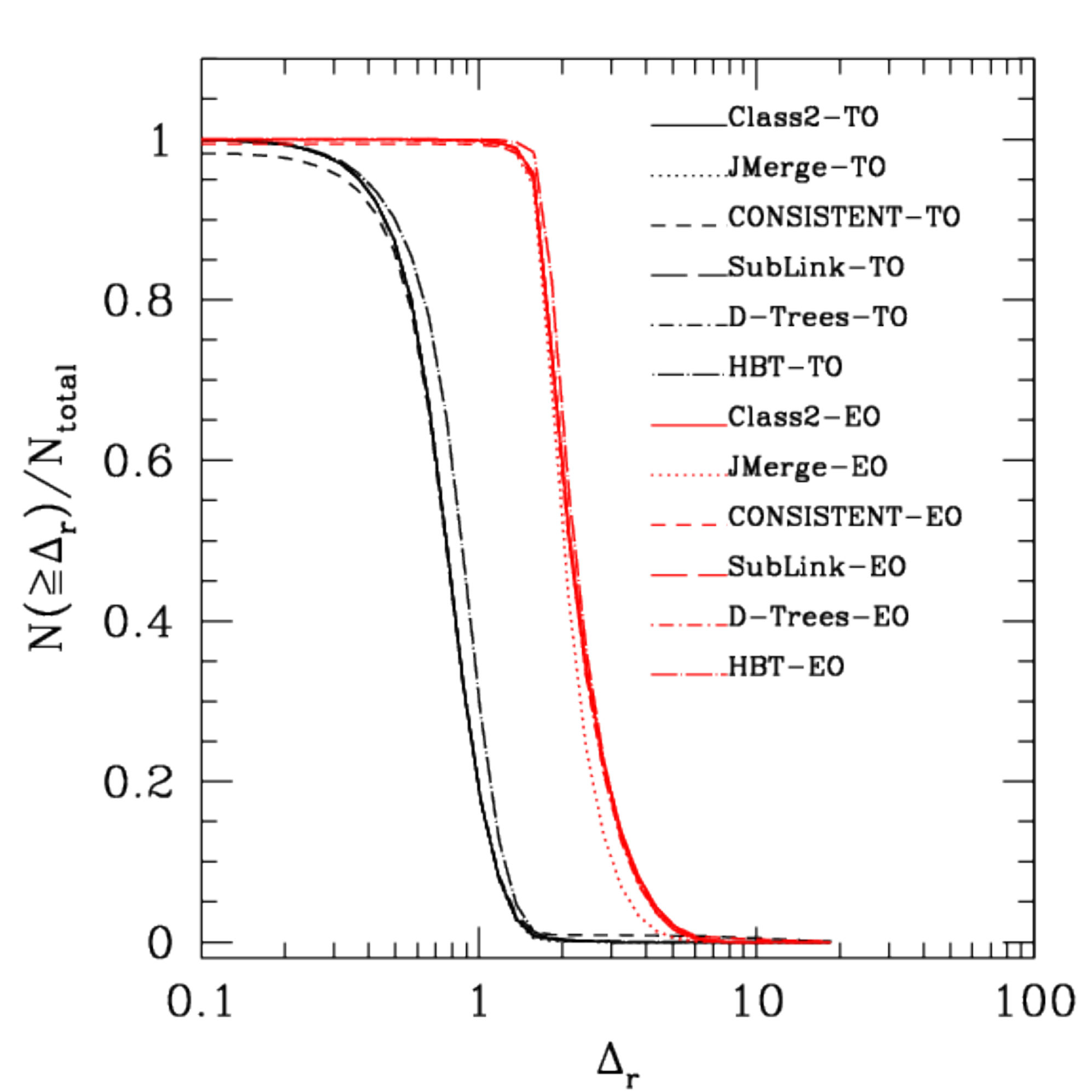}
    \includegraphics[width=0.85\linewidth]{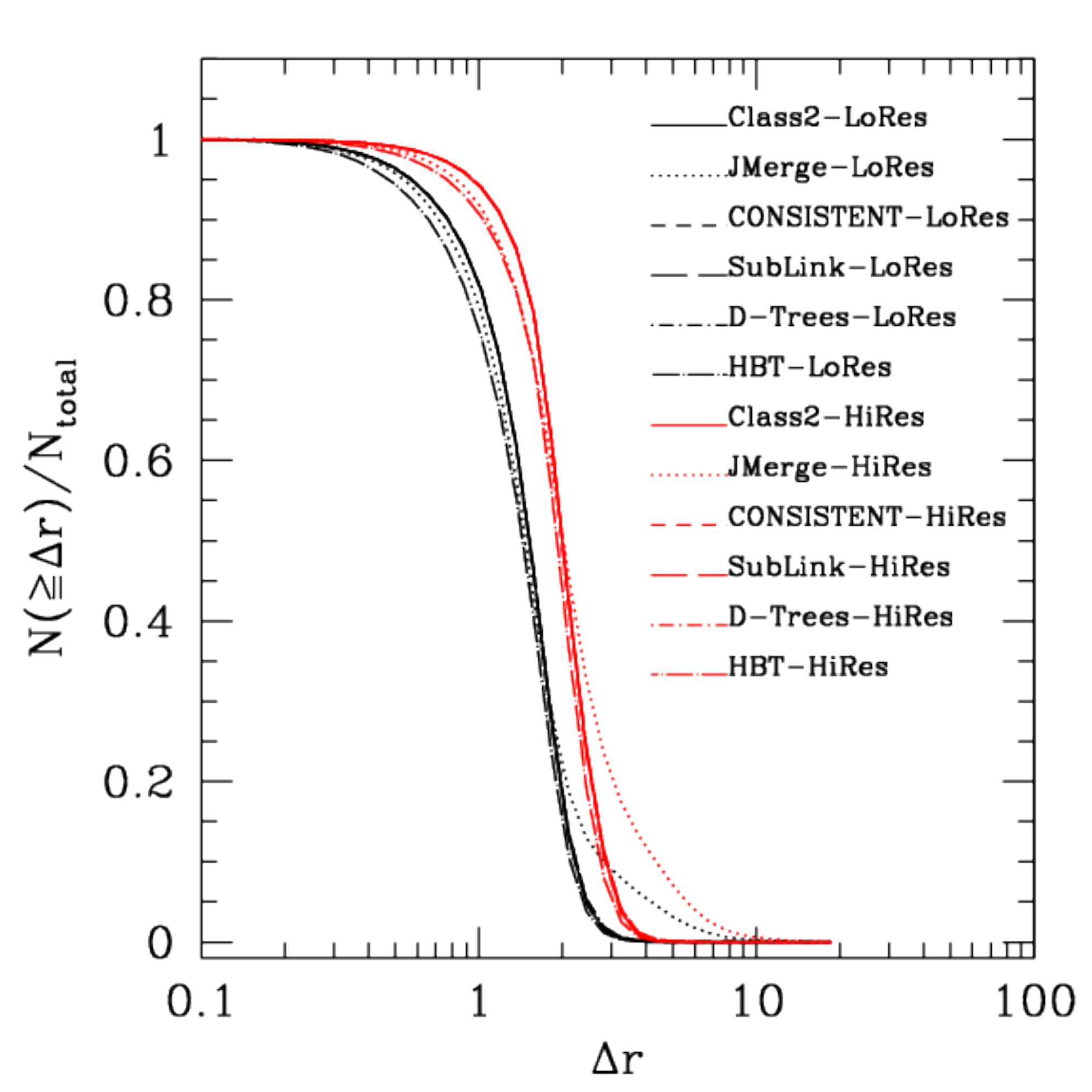}
    \caption{The displacement statistic, $\Delta r$ (see text), for
      all \halos\ and their main progenitors for which both have more
      than 200 particles. The upper panel displays results from the
      \singlehalo\ simulation. For brevity only the \TO\ (black lines)
      and \EO\ (red lines) output strategies are displayed. The lower panel displays results for the
      convergence study. Black lines are from the \lowres\ simulation
	  and red lines are from \hires\ simulation. In both panels the different styles of
	line represents different tree-builders.}
	\label{pic:mis}
\end{figure}
Before discussing the properties of merger trees, we want to ensure
that every tree builder works appropriately for the different output strategies
and mass resolutions. Here we use the following merit function, known as the
\textit{displacement statistic}, to quantify the performance of merger trees:
    \begin{equation}
    \Delta_{r}=\frac{|{\bf r}_B-{\bf r}_A-0.5({\bf v}_A+{\bf v}_B)(t_B-t_A)|}{0.5(R_{200A}+R_{200B}+|{\bf v}_A+{\bf v}_B|(t_B-t_A))} .
    \label{eq:misplace}
    \end{equation}
Here subscripts $A$ \& $B$ refer to the two snapshots being compared,
$t$ is the cosmic time, ${\bf v}$ and ${\bf r}$ are the velocity and
position of the considered (sub)halo and its progenitor, and
$R_{200}$ is the radius that encloses an overdensity of 200 times the
critical density. \cite{Srisawat13} used this formula to quantify how
far \halos\ are displaced from their expected locations when moving
from one snapshot to the next. Large values indicate a halo mismatch
in the tree. It should be mentioned that \cite{Srisawat13} only employed this
statistic for the deviation of main \halos\ because it is hard to
predict the motion of sub\halos. In this work we simply compare the
values of $\Delta_r$ arising from the various tree-building codes and
therefore we also include sub\halos in our analysis.

As the upper panel of \Fig{pic:mis} shows, most lines from the same
output strategy (with the same colour but different line style)
overlap, although the value of the turnover, which corresponds to the
peak in Figure 6 of \cite{Srisawat13} is larger. This is due to the
inclusion of sub\halos\, whose locations are harder to predict.  The
result for \hbt\ (long dash-dotted line) is to some extent different from
the others, because \hbt\ alters the underlying halo catalogue. We have verified
for all output strategies that the distributions from different
builders also agree well. For brevity only lines from \TO\ and \EO\ are
shown in the figure.

The lower panel of \Fig{pic:mis} shows the same statistic, $\Delta_r$,
for both the \hires\ and \lowres\ simulations. As we saw previously,
lines for the different codes overlap in both simulations (except
\jmerger, the dotted line). As discussed in \cite{Srisawat13},
\jmerger\ occasionally makes incorrect halo matches due to the lack of
particle ID information. The difference between the \lowres\ and
\hires\ result can be attributed to the very large difference in the
number of sub\halos. The \hires\ simulation has more sub\halos\ than
the \lowres\ simulation which makes $\Delta_r$ larger.

We conclude that varying the output strategy (even dramatically in the
case of the \LO\ dataset) or changing the mass resolution does not
break any of the contributed merger-tree-building routines and that
they all produce results in line with our expectations.

\section{Result II: Stability of Merger Tree}
\subsection{Geometry} \label{sec:S_G}
\begin{figure*}
    \centering
    \includegraphics[width=0.88\linewidth]{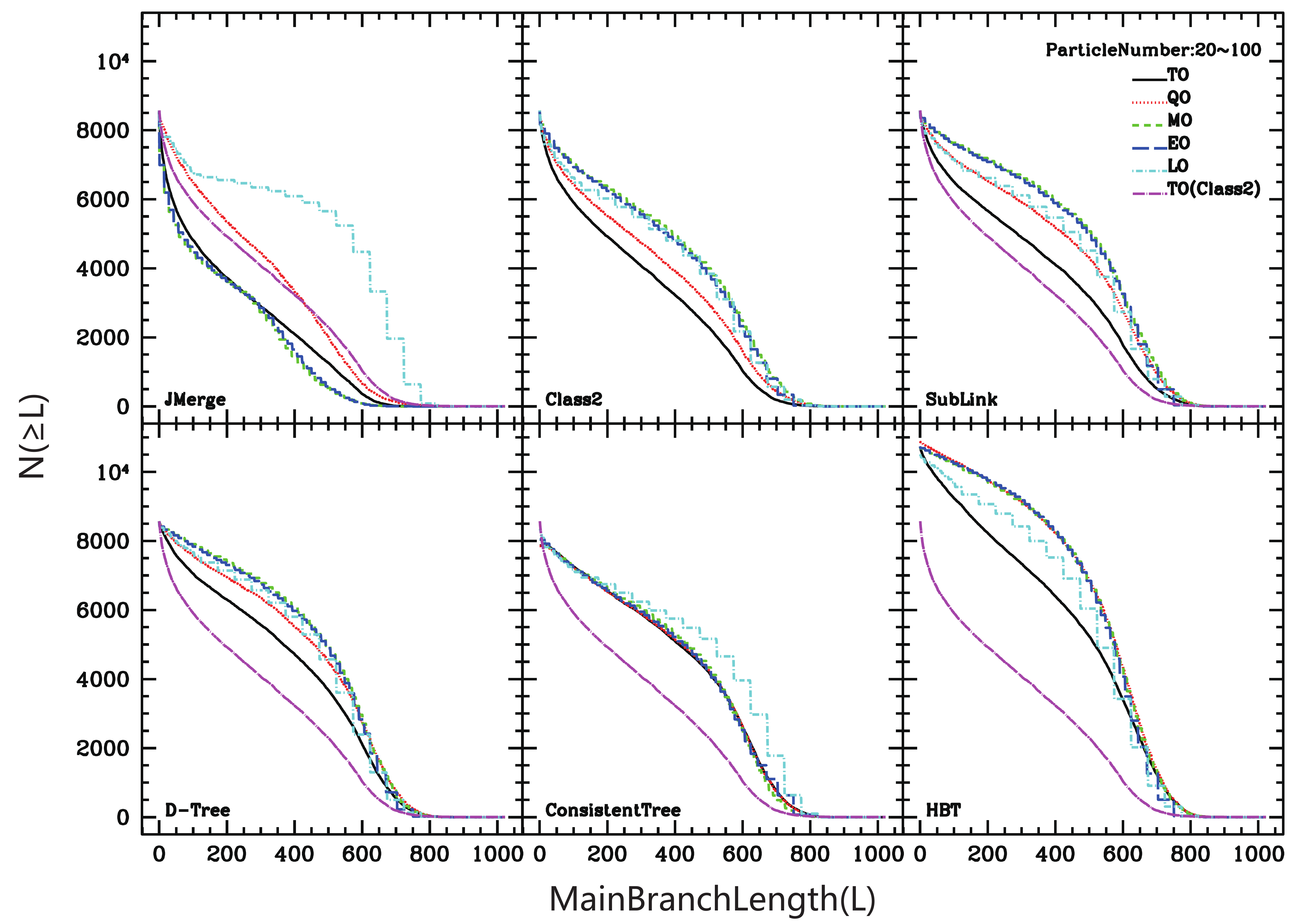}\\
    \includegraphics[width=0.88\linewidth]{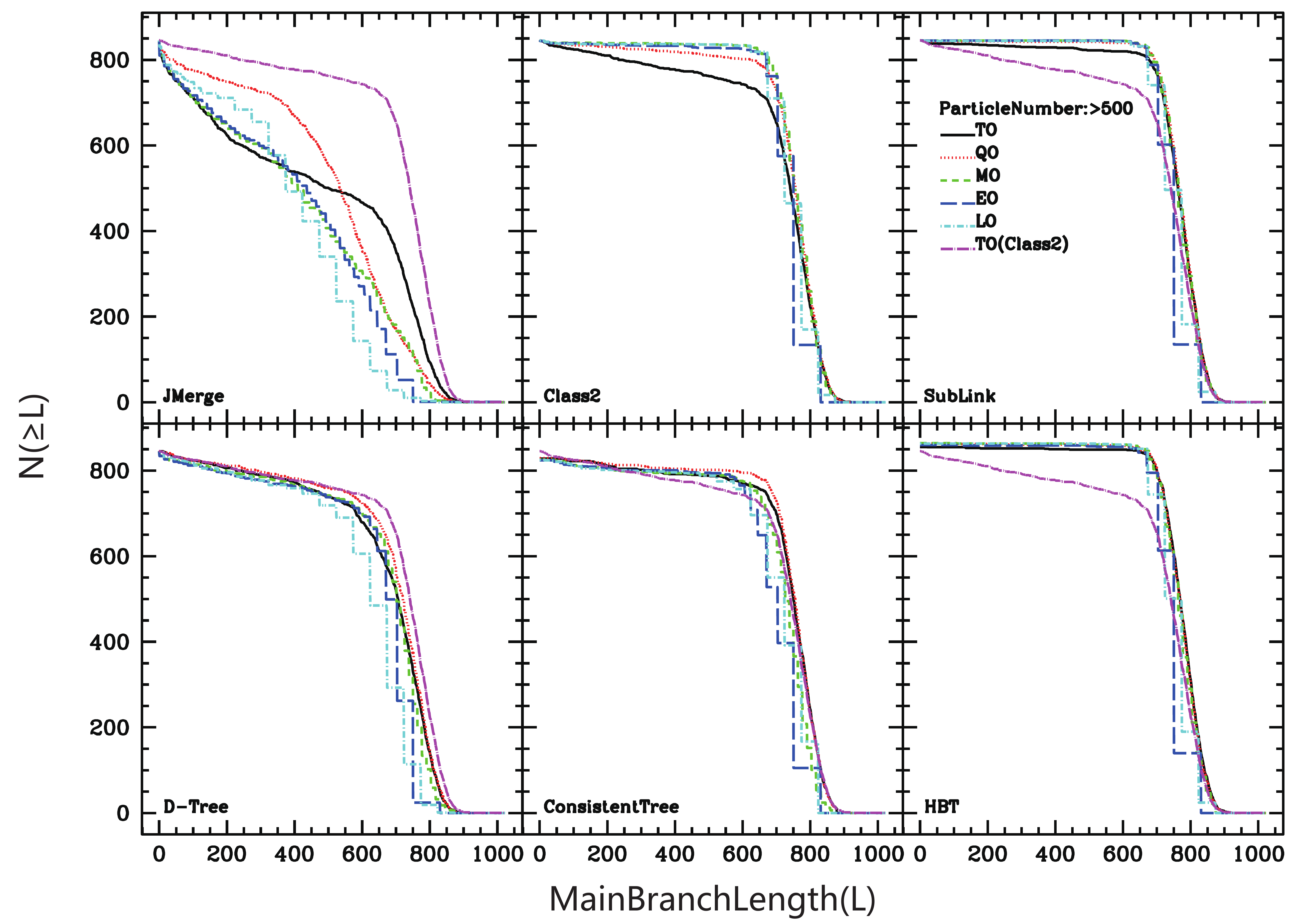}
    \caption{Cumulative number of merger trees with main branch
      lengths larger than {\it L} in the \singlehalo\ simulation. The
      upper panel with 6 subplots shows the results from merger trees
      with root \halos\ containing 20-100 particles, and the lower
      panel is for root \halos\ larger than 500 particles. Each
      subpanel shows the results of a tree-builder as labeled in the
      bottom left corner. Different line colours and styles represent
      different output strategies as indicated in the legend. The
      \TO\ line of \class\ is reproduced in magenta on all panels for
      guidance.}
    \label{pic:L:S}
\end{figure*}
\Fig{pic:L:S} shows the cumulative main branch length function for the
various output datasets as indicated in each panel. We examine merger
trees of root \halos\ with 20-100 particles, 100-500 particles and more
than 500 particles. For brevity, we only display the results for
\halos\ with 20-100 particles and more than 500 particles here. The
results from \halos\ with 100 to 500 particles are close to those from
\halos\ with more than 500 particles. For class 2 tree-builders
without patching, such as \mergertree, \treemaker, \velociraptor, and
\ysamtm, the results are almost indistinguishable, so we plot them on
a single subplot labelled \class.

For the thousand output dataset (\TO), most codes find shorter main
branches, hence producing lower curves on the plots.
This is due to additional snapshots increasing the probability of cutting
a link. The rising trend of the curves from the quarter output (\QO),
equal timed output (\EO) and Millennium output (\MO) supports this
conclusion. This effect is very clear for small \halos, and for
large \halos\ when class 2 tree-builders without patching are applied.
Patching the merger tree by using information from additional
snapshots clearly alleviates this problem to a greater or lesser
extent (we show three examples of class 3 builders that achieve
this). Of the three shown here, \consistenttree\ is the most stable to
changes in the output strategy, with almost overlapping results. Both
\sublink\ and \dtree\ could probably be adapted to achieve this too.
However, by default they interpolate over a (small) fixed number of
snapshots rather than over a fixed time-scale. For both of them the
number of patching snapshots needs to be increased in proportion to
the number of outputs, something that was not done here.

The class 4 builder, \hbt, also shows stable performance for merger trees with large root \halos. It
finds very long main branch lengths and the curves of all output
datasets overlap when building trees for \halos\ with more than 100
particles. However, it finds a somewhat low \TO\ line for halos with
$20$-$100$ particles. We suggest that the additional snapshots
increase the probability of cutting a link, because some
\halos\ may not easily be resolved by the halo finder. A halo could
be unresolvable if it falls into a larger halo, or if its mass
fluctuates taking it below the minimum particle number threshold of
the halo finder. Class 3 builders can handle this situation by
patching over the unresolved halo. However, \hbt\ does not do any
patching. It finds descendants just in the next snapshot. Its unique
algorithm allows it to track sub\halos\ accurately but it can still
lose track when a main halo fluctuates to too a small size. The latter
situation happens more for main \halos\ with a small number of particles.

In all the panels the curves of \MO\ and \EO\ are generally almost
indistinguishable. This implies that small differences in the interval
between outputs do not affect the merger tree too much, especially
when such differences occur in the early stages of the simulation.

The lame output strategy \LO\ causes problems for the class 1 tree
builder because wildly changing time intervals between snapshots makes
extrapolation of halo positions to enable matching very difficult. The
\LO\ dataset does not cause too many problems for the class 2 tree
builders, although in some cases the \LO\ lines are slightly lower than
those of the \EO\ and \MO\ datasets. For small \halos\ with the class
3 patching tree builders, this difference becomes clearer. This implies
that the unconventional output strategy \LO\ may prevent the patching process from
optimizing the merger tree.

\begin{figure}
    \centering
    \includegraphics[width=1\linewidth]{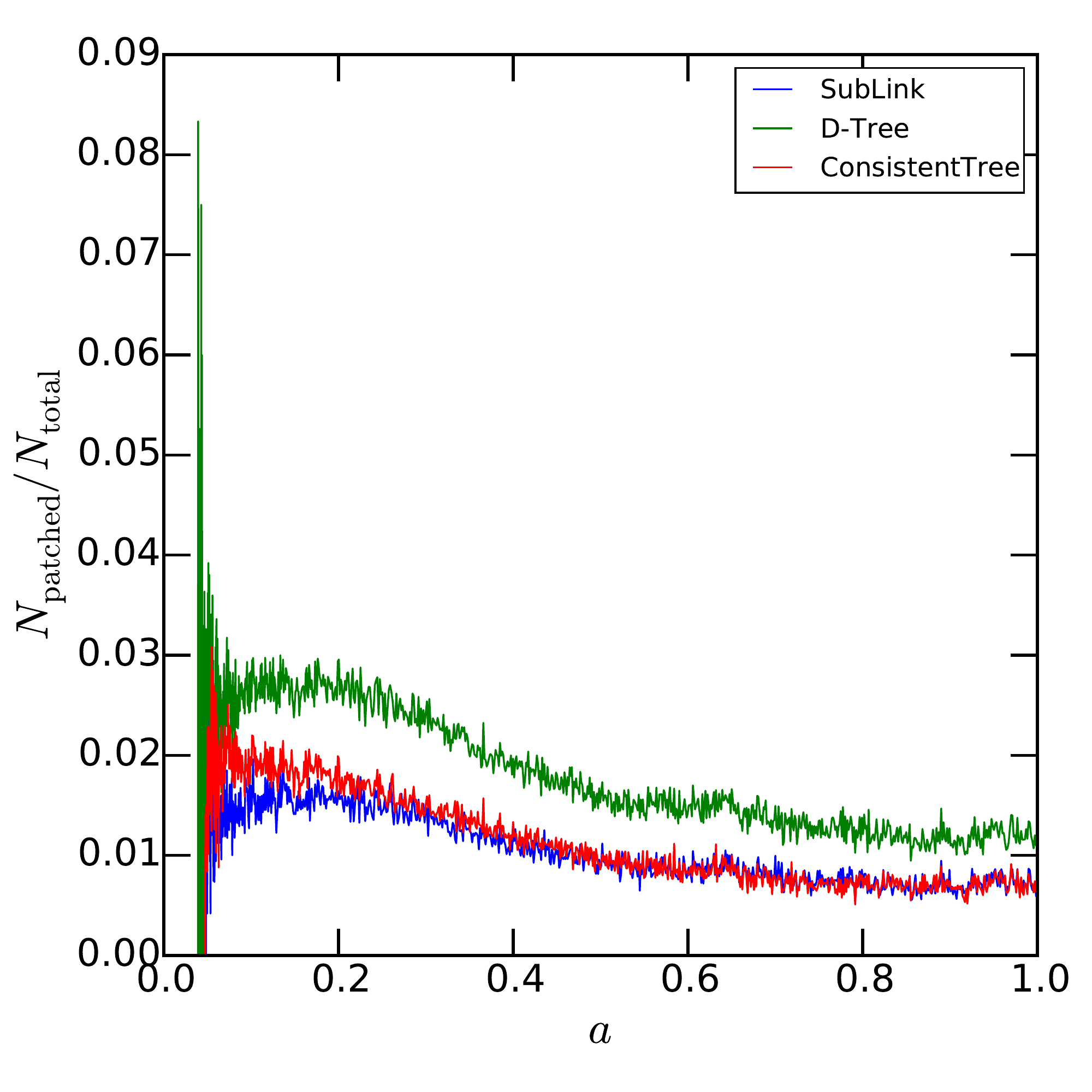}
	\caption{The fraction of patched \halos\ in every snapshot in the \TO\
	dataset as a function of the scale factor, $a$. Each line represents a
different \textsc{Class}3 tree builder as indicated.}
    \label{pic:missed}
\end{figure}

Using our \textsc{Class}3 routines we can identify how many \halos\ are
being missed in the \class\ methods by noting when the \textsc{Class}3
methods patch the tree by inserting an extra halo. We show this in
\Fig{pic:missed}, which gives the fraction of patched \halos\ for
\sublink, \dtree\ \& \consistenttree\ for the \TO\ output set. All \halos\
are taken into account in this figure. We see
that for all three methods at least 1 percent of the \halos\ are required
to be patched at all times and that this fraction rises at early
times. Conversely, without patching, at least 1 percent of the \halos\ are
missed from the trees at all times. We have tested the dependency
of this missing fraction with the number of snapshots used. From 1 percent
for the \TO, the
fraction rises slowly as the number of snapshots is reduced, roughly
doubling for the 64 snapshot strategies, i.e. about 2 percent of
\halos\ are missed by the 64 output strategies, \MO\ \& \EO\ at
$z=0$.

\begin{figure*}
    \centering
    \includegraphics[width=0.8\linewidth]{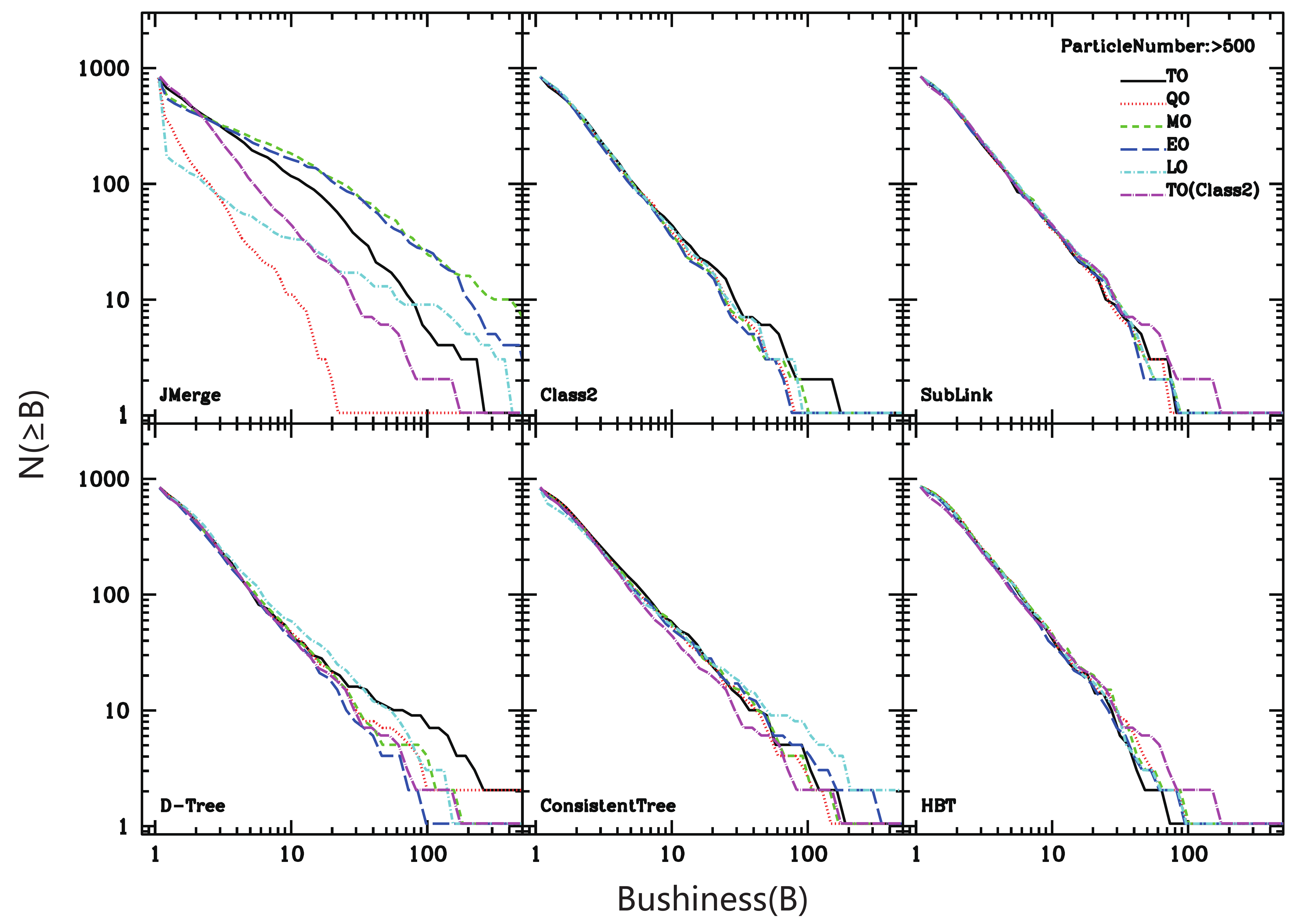}
    \caption{Merger tree bushiness $B$ for each of the builders in \singlehalo\ simulation. Only
      root \halos\ with more than 500 particles are selected
      here. Each subpanel shows results for an individual tree-builder as indicated,
      except for the four class 2 finders which are indistinguishable
      and are therefore all shown on the same subpanel. Different line
      colours and styles represent different output strategies as
      indicated in the legend. The \class\ \TO\ line is reproduced on
      all panels for guidance.}
    \label{pic:B:S}
    \centering
    \includegraphics[width=0.8\linewidth]{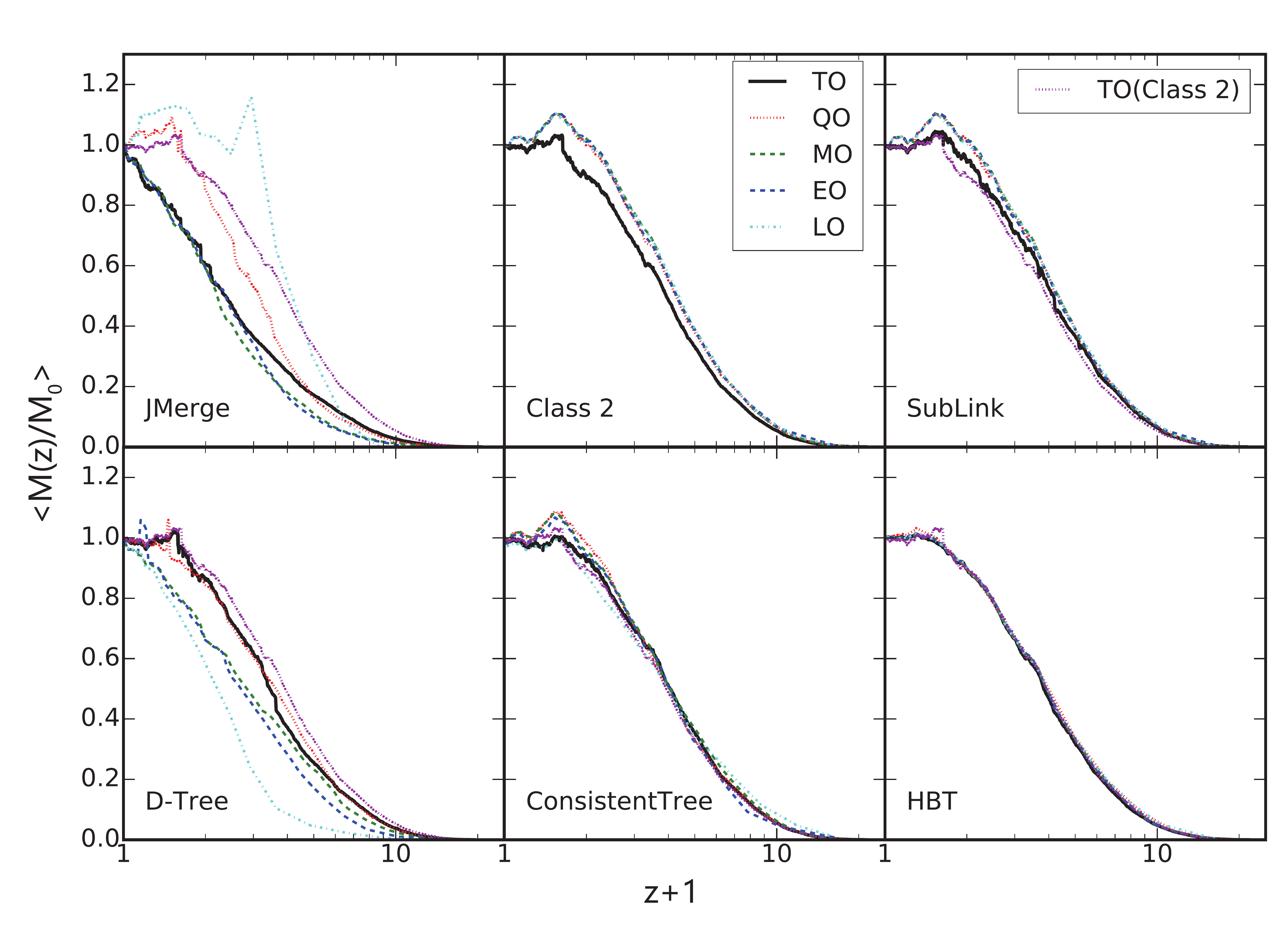}
    \caption{Average mass history for each of the builders as a
      function of redshift. This corresponds to the total mass in tree main branches for root
      \halos\ between $0.5 \times 10^9 {\rm M_\odot}$ and $1.5 \times
      10^9 {\rm M_\odot}$ at $z=0$, normalized by the mass at
      $z=0$. Different lines represent merger trees built from
      different output strategies, as indicated by the legend.  Each
      subpanel displays results for one tree builder as indicated,
      except for the four class 2 finders which are indistinguishable
      and so are all shown on the same subpanel. The \TO\ line of
      \class\ is reproduced in magenta on all panels for guidance. }
	\label{pic:mr}
\end{figure*}

\Fig{pic:B:S} shows that, with the exception of \jmerger, the
bushiness of the trees does not change for different output
strategies. We only show trees with root \halos\ larger than 500
particles here, since the results for the other ranges are
similar. This result indicates that the merger history is rarely
affected by the output strategy and that in terms of bushiness all the
class 2, 3 \& 4 merger-tree-builders produce trees that look very
similar. The exception is \jmerger\ which produces both significantly
more very bushy trees (i.e. trees with very short main branches
compared to the number of branches), as well as significantly more
`bare' trees that consist almost entirely of the main branch. These
differences can again be attributed to the difficulty in matching
\halos\ between snapshots when no particle ID information is
available, a consequence of which is oddly shaped trees.

\begin{figure*}
    \centering
    \includegraphics[width=1\linewidth]{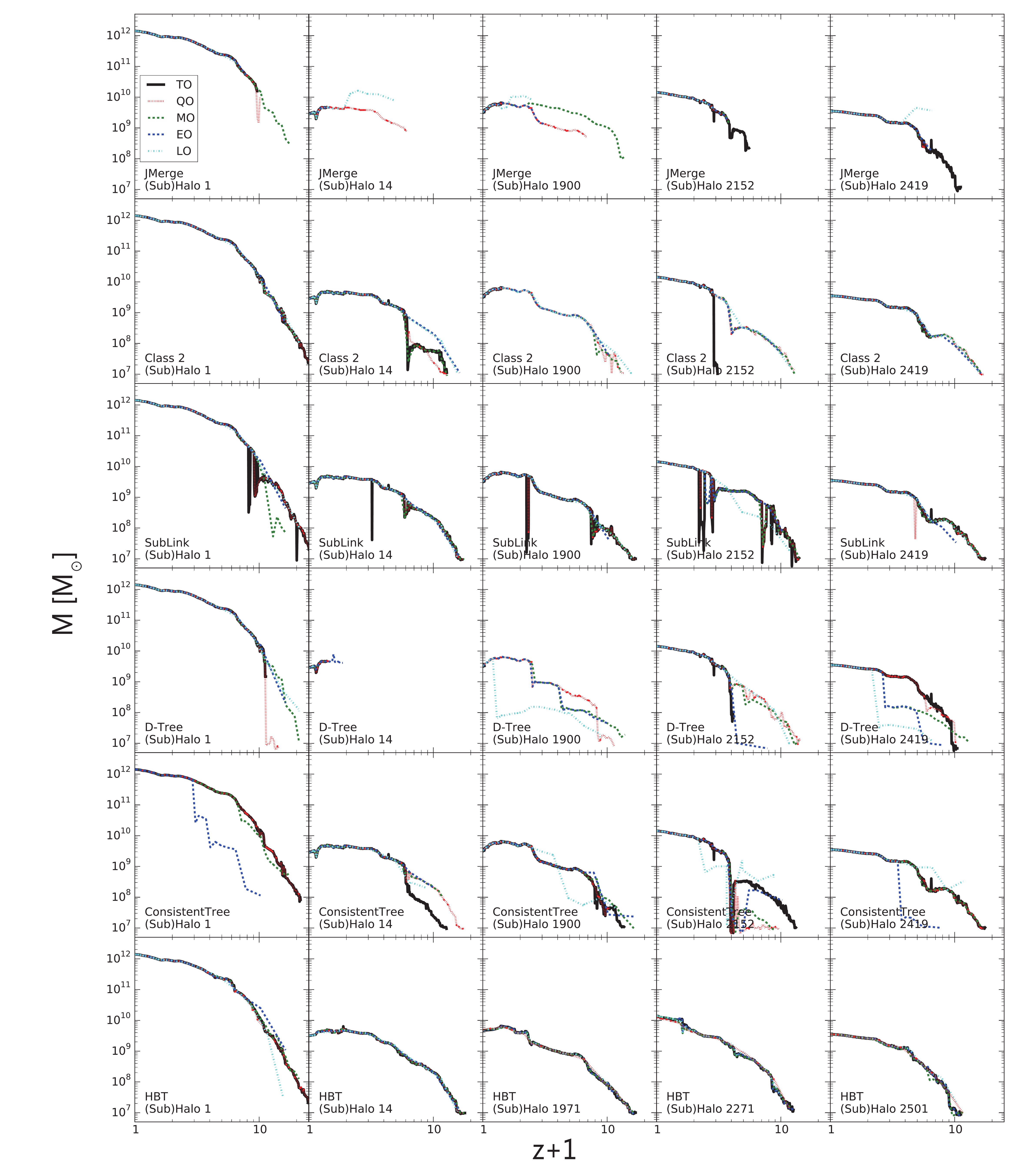}
    \caption{Mass histories for 5 (sub)\halos\ constructed by
      different merger-tree-builders. Different line styles represent
      different output strategies, as shown by the legend.  Each row
      shows results for one tree builder as indicated, except for the
      four \class\ finders which are indistinguishable and so are all
      shown on the second row.}
	\label{pic:mah}
\end{figure*}

\begin{figure*}
    \centering
    \includegraphics[width=1\linewidth]{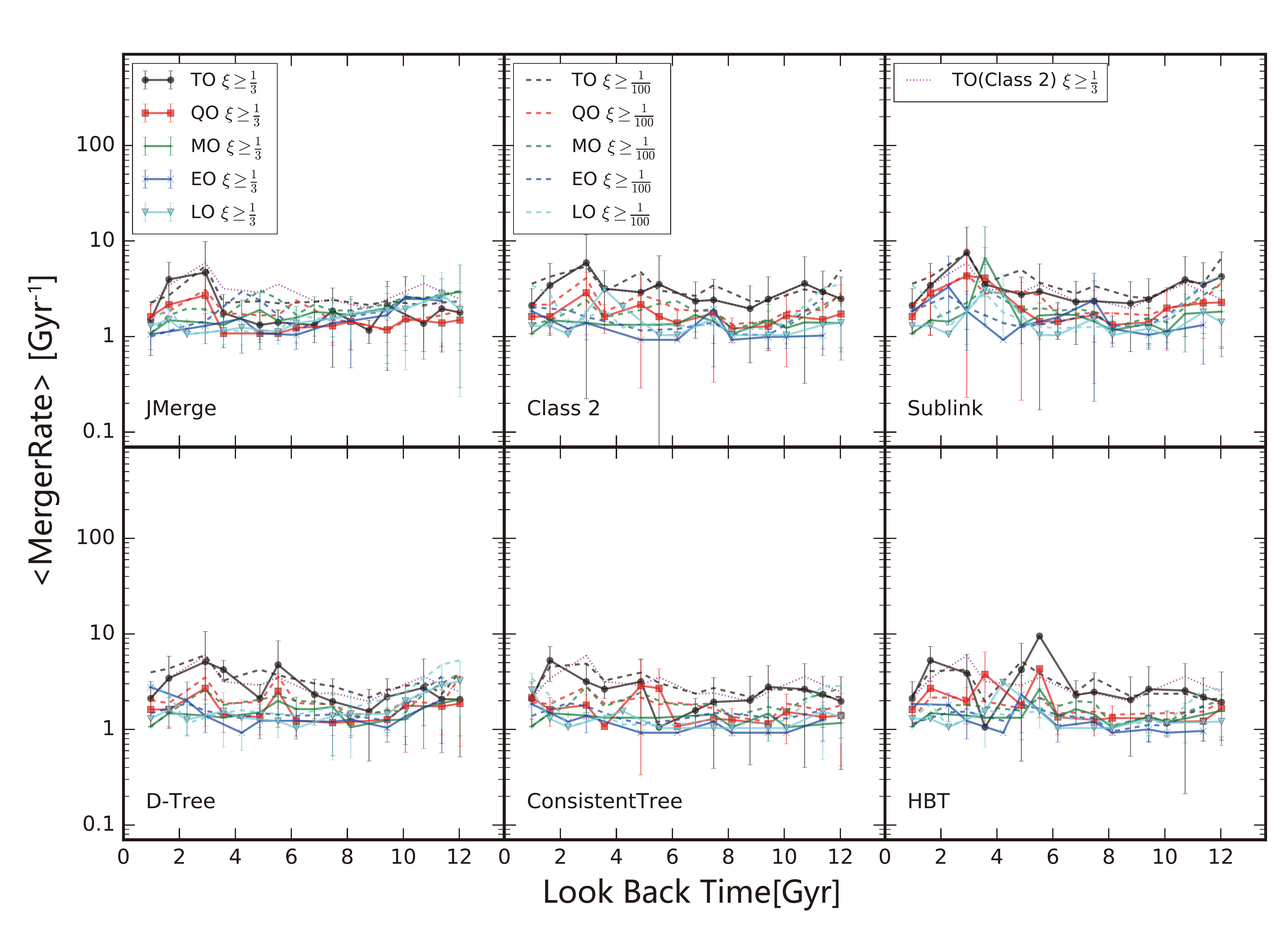}
    \caption{Mean merger rate (mergers per descendant halo per $\rm
      Gyr$) as function of look-back time.  Lines with different
      colour represent the merger tree built with a different output
      strategy, as shown by the legend. Solid lines show the merger rate
      with progenitor mass ratio larger than $1/3$. Dashed lines show
      the merger rate with progenitor mass ratio larger than $1/100$.
      The \TO\ line of \class\ is reproduced in magenta on all panels
      for guidance. The error bars show the rms in every time bin.
      Each subpanel shows results for one tree builder as indicated,
      except for the four class 2 finders which are indistinguishable
      and so are all shown on the same subpanel.}
	\label{pic:merger}
\end{figure*}
\subsection{Mass history}
\Fig{pic:mr} illustrates the mass history for
\halos\ between $0.5 \times 10^9 {\rm M_\odot}$ and $1.5 \times 10^9 {\rm
M_\odot}$ at $z=0$, normalized by the mass contained within this set at $z=0$.
The different line styles indicate the different output sets as detailed in
the legend. This confirms the results seen earlier for main branch length
which indicated that most of the builders find less material within \halos\
for the \TO\ output set compared to the sets with fewer outputs. \jmerger\
again struggles in comparison to all the builders that use particle IDs, with
a dramatic growth in \halos\ at late times due to the very truncated trees
this method often produces. \hbt\ produces extremely stable results by this
measure, as was indicated to some extent by \Fig{pic:L:S}.

We study individual tree mass histories in \Fig{pic:mah}. In
this figure we compare 5 \halos' merger trees across the different
output sets and tree builders. Each column shows one halo's mass
accretion history and each row shows the merger trees constructed by
the same tree builder as indicated. The matched \hbt\ (sub)\halos'
have different IDs because \hbt\ constructs its own halo catalogue.
Different line styles represent different output sets as indicated on
the legend. The lines from the \TO\ set are in bold so that the
\TO\ output can easily be distinguished from the other sets. In row 2,
which shows the mass growth of merger trees built by \class\ tree
builders, we see several examples of the cutting of the main branch in
the \TO\ output set.

Rapid declines in mass and violent fluctuations are also seen for many
tree builders. These typically arise due to mergers, where
distinguishing \halos\ and assigning masses becomes difficult, as reported by
\citet{Srisawat13} and \citet{Avila_2014}. \citet{Behroozi2015}
explore this issue further by studying the consequences of major
merger events.

\Fig{pic:mah} can also help us understand how to implement
patching algorithms which aim to bridge over the drop-outs particularly
seen for the \class\ builders. Such patching may also aim to smooth
over the rapid and unphysical mass fluctuations seen here.  As can be
seen in \Fig{pic:mah} the class 3 tree builders we have tested all
work to some extent, although some issues remain for some \halos.  For
example for halo 14 we see an illustration of the re-appearance of a
halo for the \dtree\ finder (second column).  \dtree\ also has issues
with halo 1900 (third column). We stress that we could have found
similar examples for all the class 3 builders and that such instances
are far less common for the class 3 builders compared to \class.  As
expected the bottom row of \Fig{pic:mah}, which shows merger trees
built by \hbt, is relatively stable. \hbt\ constructed its own
(sub)halo catalogue during the process of building its merger
trees. This extra work suggests that a good starting halo catalogue is
an important factor in constructing merger trees. For more information
about comparing the influence of the input halo catalogue, see
\citet{Avila_2014}.

\subsection{Merger rate}

\Fig{pic:merger} shows the mean merger rate as function of look-back time (in
units of ${\rm Gyr}$). \Halos\ with $z=0$ masses between
$(1\pm0.5)\times10^9{\rm M_{\odot}}$ are selected. When calculating the merger
rate, the number of merger events is divided by the time interval over which
this number of mergers are seen. Because the time interval between snapshots
can be quite small, this can lead to unstable merger rate. In this case multiple
snapshots are aggregated.  For consistency, the time interval is fixed. In the
\EO\ output set, it is $0.81 {\rm Gyr}$. In the other sets, the time interval
varies from $0.8 {\rm Gyr}$ to $1.05 {\rm Gyr}$, because the snapshots cannot
be exactly matched.

As for our earlier results for bushiness, in \Fig{pic:merger} we show
that the mean merger rate for each output set is essentially
indistinguishable. Given the similarity between these two measures
this is not entirely surprising. It is interesting that the Class 3
and Class 4 finders that include patching are not significantly
different from the \class\ finders. In practice mergers are readily
detectable and although some 'flyby' events may be misclassified as
mergers, the number of such events is small.

The merger rate shown in \Fig{pic:merger} is different from Figure 8 in
\citet{Fakhouri2008}. This is because in \singlehalo\ most \halos\ are
sub\halos, \Fig{pic:merger} actually shows the merger rate of
sub\halos.

\section{Result III: Convergence of Merger Tree}
\begin{figure*}
    \centering
    \includegraphics[width=0.8\linewidth]{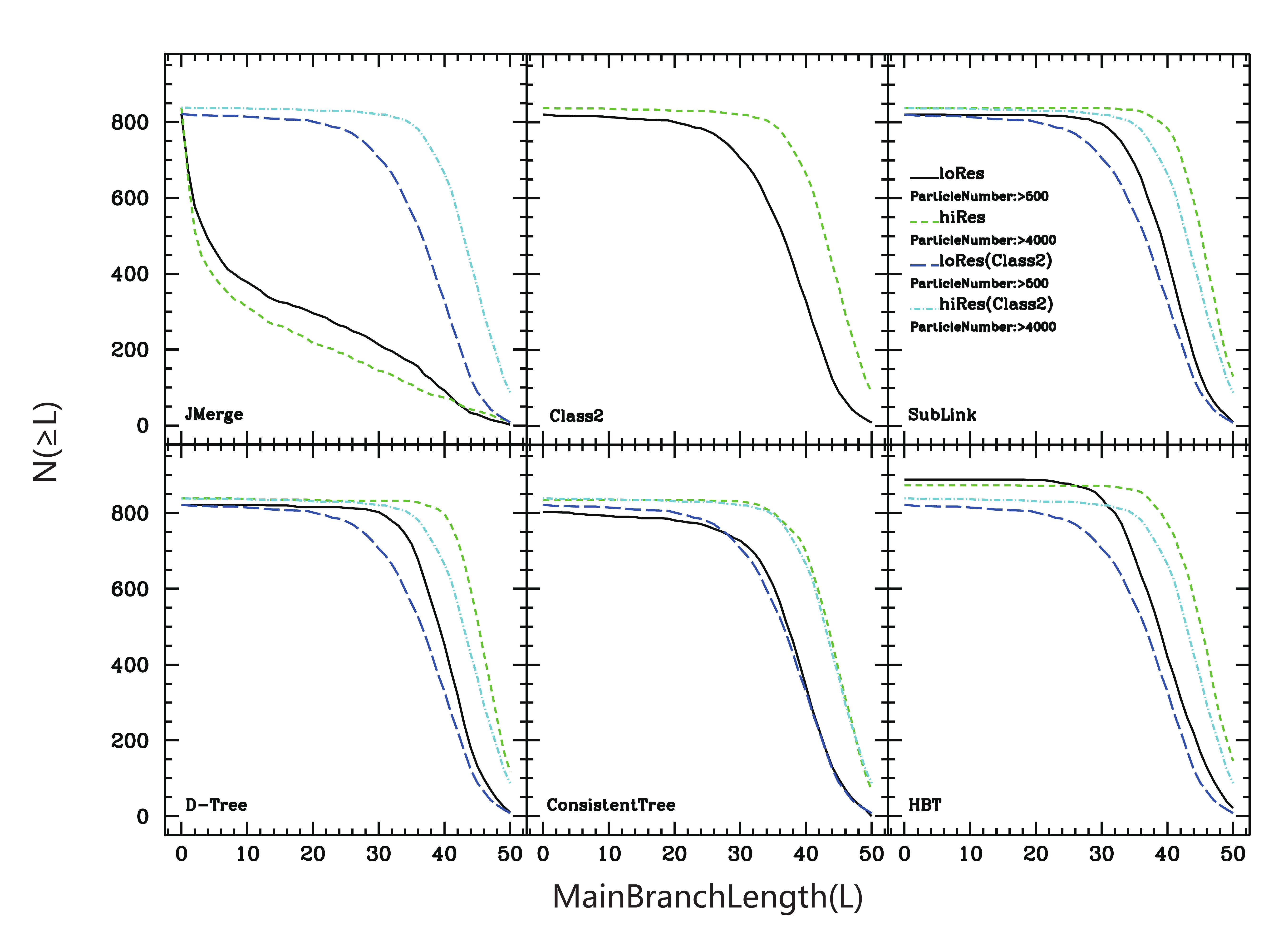}
    \caption{Cumulative number of trees with main branch lengths
      larger than {\it L} in the \lowres\ and
      \hires\ simulations. Root \halos\ with 20-100 particles in the
      \lowres\ simulation are selected. Trees with root \halos\ with
      the same particle number or the same halo mass in the
      \hires\ simulation are selected respectively. Each panel
      represents one tree builder as indicated except for the four
      class 2 builders which are shown together on the
      \class\ panel. Different line types and colours represent
      different datasets as indicated in the legend.The
      \lowres\ and \hires\ lines of \class\ is reproduced in blue and cyan on all panels for
      guidance.}
    \label{pic:Convergence:L:1}

    \centering
    \includegraphics[width=0.8\linewidth]{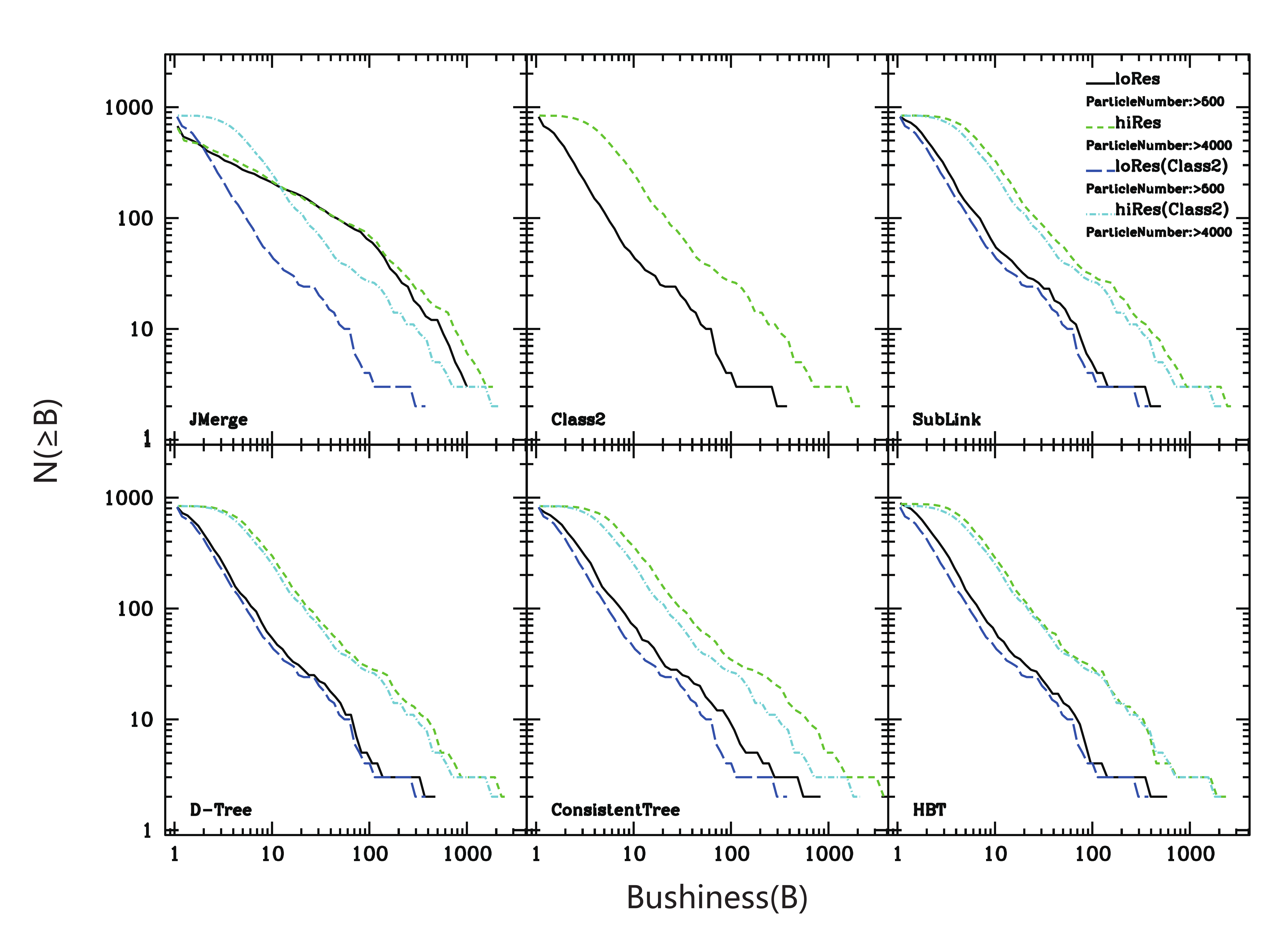}
    \caption{Number of trees of each bushiness {\it B} in the \lowres\ and \hires\ simulations. The criterion
      for selection and the linetypes are the same as for
      \Fig{pic:Convergence:L:1}}
    \label{pic:Convergence:S:1}
\end{figure*}
\subsection{Geometry}

\Fig{pic:Convergence:L:1} shows the cumulative number of trees with
main branch lengths larger than {\it L} in the \lowres\ and
\hires\ simulations. To investigate the influence of resolution, we
draw two lines for comparison: trees with root \halos\ containing more
than 500 particles in the \lowres\ box (black solid line), and
equivalent trees with root \halos\ containing more than 4000 particles
in the \hires\ box (green dashed line). This choice corresponds to
\halos\ more massive than $10^9{\rm M_{\odot}}$ in both
simulations. The green dashed line represents the same halo population
as the black solid line, simulated at higher resolution. The
comparison between the black line and the green line indicates that,
for trees with the same root halo mass range, better mass resolution
results in longer main branches. All the codes perform alike except
for \jmerger\ which appears to show the opposite trend. For trees with
smaller root \halos\ the trends are similar to
\Fig{pic:Convergence:L:1}, though the deviation between the green line
and the black line is larger. We can also see that patching class 3
and higher tree builders still find slightly longer main branches than
the class 2 builders in both simulations (the \class\ curves are
overplotted on the other panels in blue and cyan for comparison).

\Fig{pic:Convergence:S:1} shows that resolution will also affect the
bushiness of the merger trees. The peak of the green line shifts right
compared to the black line, indicating that \halos\ with the same mass
have bushier trees in the higher resolution simulation.  As seen
previously, all the codes perform alike except for \jmerger\ which is
resolution independent unlike all the other builders. In practice two
factors drive an increase in bushiness: either there is a dramatic
decrease in the main branch length or an increase in the number of
secondary branches leading to an increase in the number of merger
events. Since we have seen that the main branch length becomes larger
with increasing resolution, we can rule out the first of these. Rather,
the increased resolution leads to an increase in the number of minor
mergers and hence the measured bushiness of the trees, an effect that
outweighs the slightly longer length of the main branches.

\begin{figure}
    \centering
    \includegraphics[width=1\linewidth]{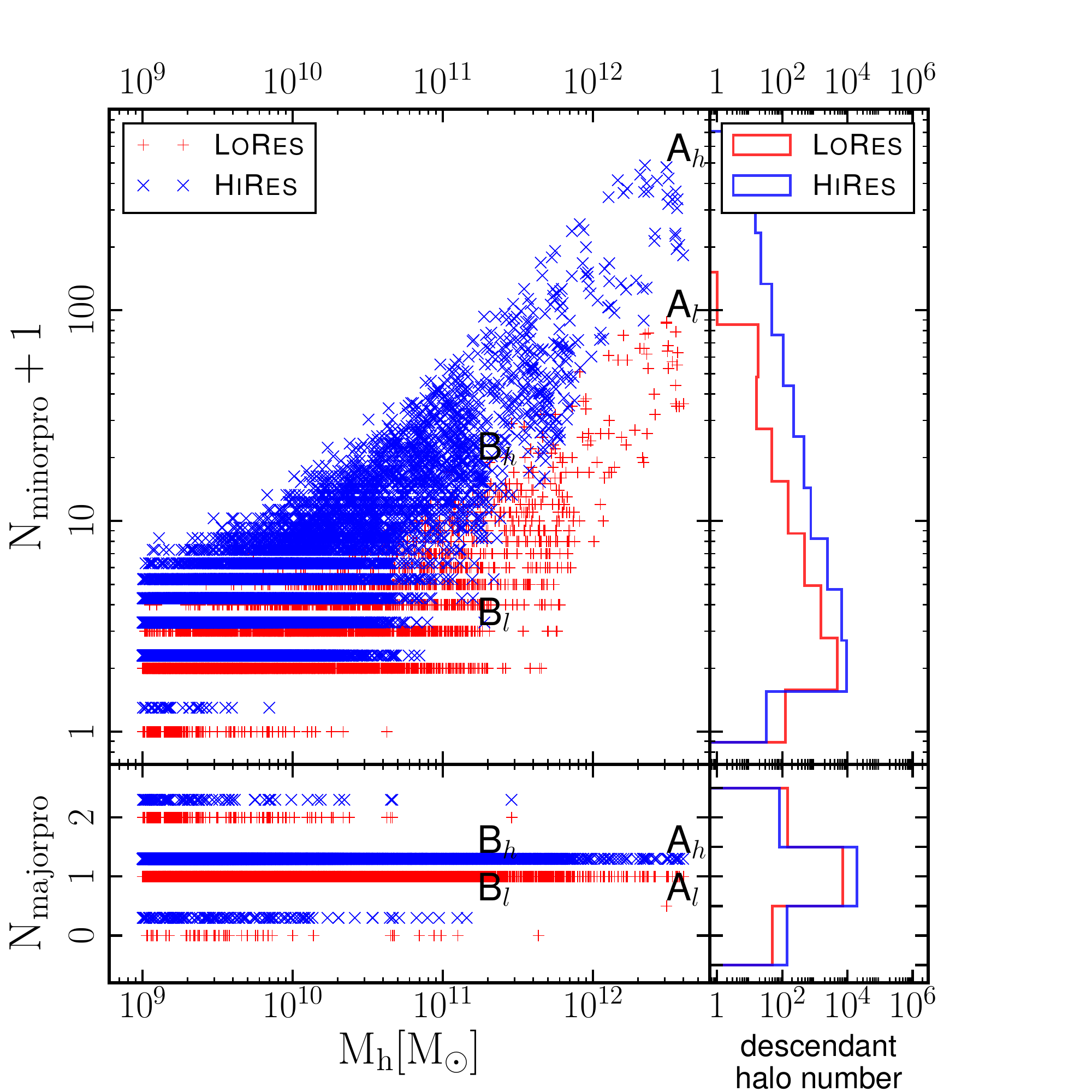}
    \caption{The number of progenitors a halo has as a function of its
      mass and histogram of the number of progenitors. The left lower
      panel counts the number of major progenitors, which have a mass
      larger than one third of their descendants', and the left upper
      panel counts the number of minor progenitors, which have a mass
      smaller than one third of their descendants'. Red pluses represent
      \halos\ from the \lowres\ simulation and green crosses represent
      \halos\ from the \hires\ simulation. The right lower panel
      represents the histogram of the number of major progenitors, and
      the right upper panel represents the histogram of the number of
      minor progenitors. Red lines indicate the \lowres\ simulation
      and green lines indicate the \hires\ simulation. All
      \halos\ from all snapshots with more than one progenitor are
      included. For clarity, points from the \hires\ simulation are
      shifted up slightly.}
    \label{pic:Convergence:P}
\end{figure}

To test this, we plot \Fig{pic:Convergence:P} to look into the details
of progenitors. We select \halos\ with more than one progenitor from
all snapshots, and plot the number of progenitors against their
mass. Since all codes (except \jmerger) look alike in
\Fig{pic:Convergence:S:1}, changes due
to the resolution make little difference among the tree builders. So
we only plot \Fig{pic:Convergence:P} for \mergertree. The images for
the other builders (except \jmerger) are very similar. The letters $A_l$,
$A_h$, $B_l$, \& $B_h$ in the figure refer to four specific \halos\ we chose to
investigate. $A_l$ and $B_l$ are \halos\ in the \lowres\ box, $A_h$ is halo $A_l$ in
the \hires\ box and $B_h$ is halo $B_l$ in the \hires\ box.

Here we introduce the terms 'major progenitor' and 'minor progenitor'
to aid the description. If a progenitor's mass is more than 33 percent of
its descendant's, we call it a major progenitor; otherwise, we call it
a minor progenitor. We separate these two kinds of progenitors in the
statistic. The lower panel of \Fig{pic:Convergence:P} shows that the
number of major progenitors does not change due to the resolution,
while the number of minor progenitors shifts to higher values in the
\hires\ box. In the subplots to the righthand side, we plot the
histogram of the numbers of major and minor progenitors in the
different resolutions. This figure gives yet another result: although the
bushiness, which is equivalent to the average number of progenitors
throughout a halo's evolutionary history, is affected by the
resolution, increasing the resolution will only increase the number of
minor progenitors.

The increase in minor progenitors is mainly due to the fact
that we can resolve smaller mass (sub)\halos\ in the higher resolution
simulation. This is a problem that chiefly concerns halo finding
itself rather than the merger-tree-builder. We compared the location
of the progenitors of halo $A_l$ and halo $A_h$, to halo $B_l$ and halo $B_h$. While
their number increases, none of these small progenitors can be matched
in position. In the non-linear regime much of the small scale
structure is totally different even if two simulations share the same
initial condition. This is a general issue beyond the scope of this
paper so we will not discuss it further here.

\subsection{Mass history}
\begin{figure*}
    \centering
    \includegraphics[width=0.8\linewidth]{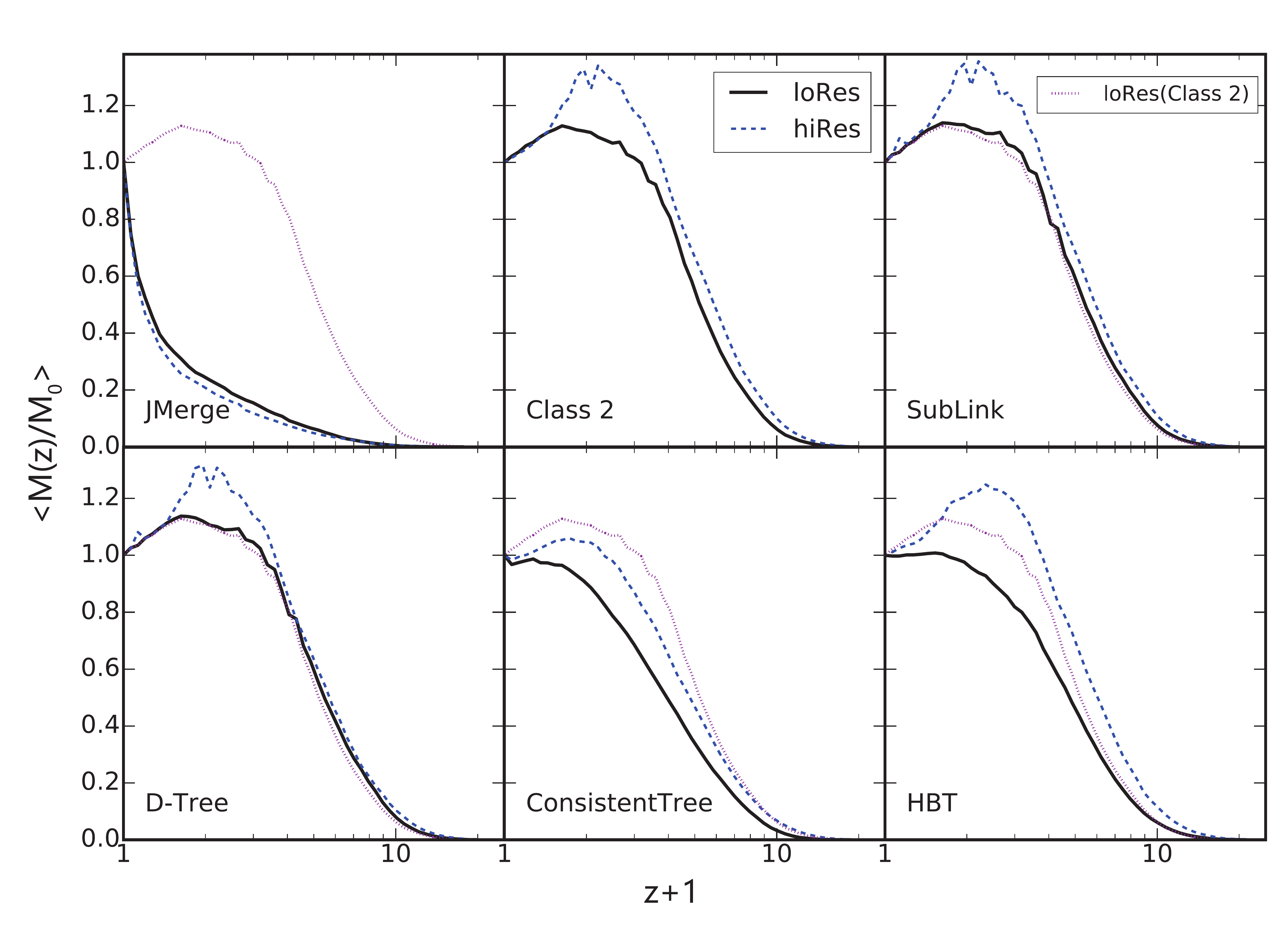}
    \caption{Average mass history for each of the builders as a
      function of redshift. Total mass in tree main branches for root
      \halos\ between $0.5 \times 10^9 {\rm M_\odot}$ and $1.5 \times
      10^9 {\rm M_\odot}$ at $z=0$, normalized by the mass at
      $z=0$. Different line styles represent the merger trees built in
      the \hires\ and \lowres\ simulations, as shown by legend.  Each
      subpanel displays results for one tree-builder as indicated,
      except for the four class 2 finders which are indistinguishable
      and so all shown on the same subpanel. The \TO\ line of
      \class\ is reproduced in magenta on all panels for guidance. }
	\label{pic:mrc}
\end{figure*}

As for \Fig{pic:mr}, \Fig{pic:mrc} shows the mass history
for \halos\ in the two simulations. Except for \jmerger, all the tree
builders find similar mass histories, with a bulge in the
\hires\ simulation due to higher mass resolution. This is consistent
with the geometry investigation.  In the \hires\ simulation, the
smaller (sub)\halos\ extend the merger trees to earlier times, which
results in a longer main branch. \jmerger\ again shows dramatic mass
accretion into trees at late times, a result of the many broken main
branches in produces.

\begin{figure*}
    \centering
    \includegraphics[width=0.8\linewidth]{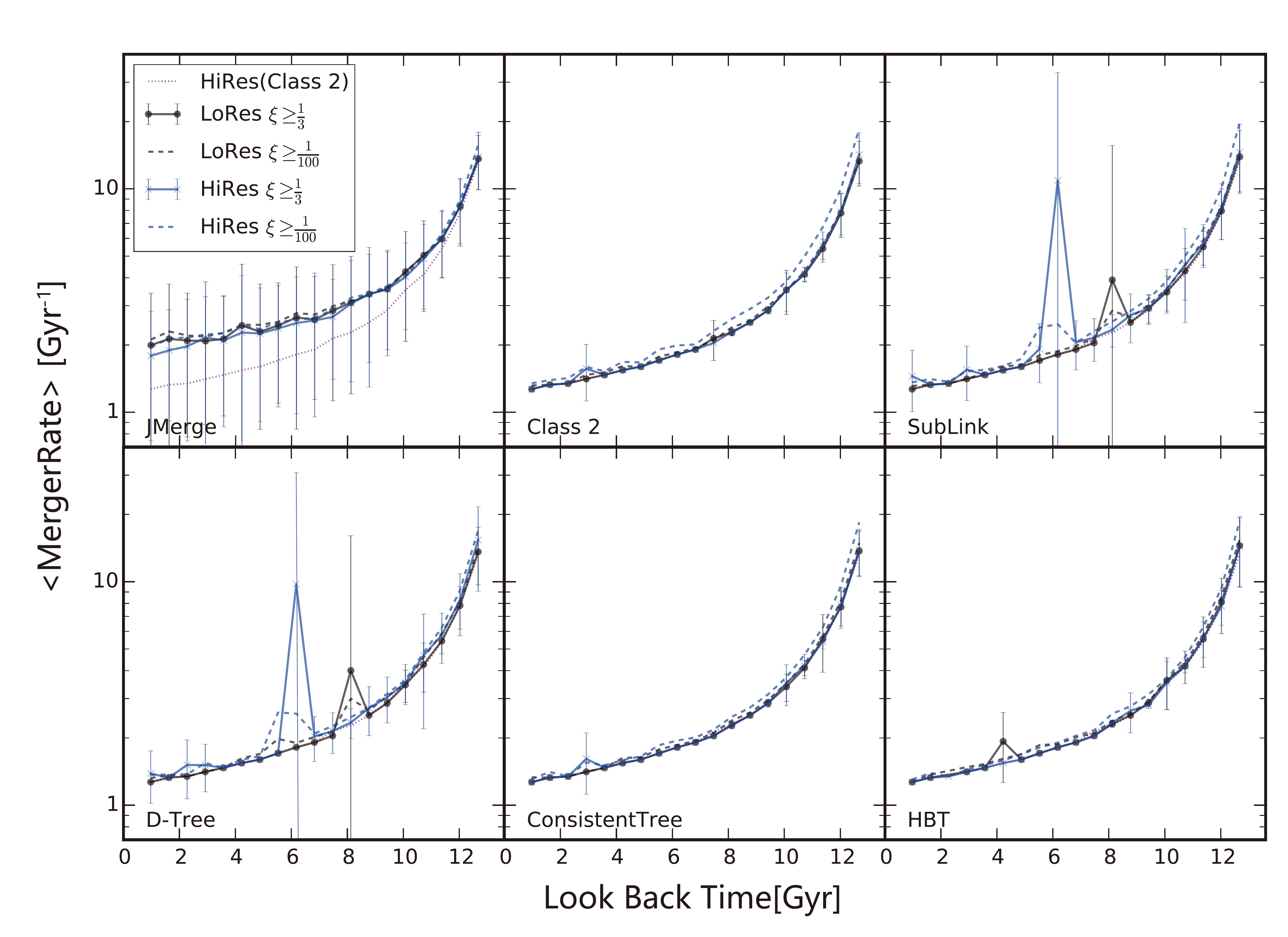}
    \caption{Mean merger rate (mergers per descendant halo per $\rm
      Gyr$) as a function of look back time.  Lines with different
      colour represent the merger trees built in the \lowres\ and
      \hires\ simulations, as shown by the legend.  Solid lines show the
      merger rate with progenitor mass ratio larger than $1/3$.
      Dashed lines show the merger rate with progenitor mass ratio
      larger than $1/100$.  The \hires\ line of \class\ ($\xi\ge1/3$)
      is reproduced in magenta on all panels for guidance.  The error
      bars show the rms in every time bin.  Each subpanel displays
      results for one tree-builder as indicated, except for the four
      \class\ finders which are indistinguishable and so all shown on
      the same subpanel.  }
	\label{pic:mergerc}
\end{figure*}

\subsection{Merger rates}

\Fig{pic:mergerc} shows the mean merger rate in the \lowres\ and
\hires\ simulations. \Halos\ with mass in the range of
$(1\pm0.5)\times10^9{\rm M_{\odot}}$ are selected.  All the tree
builders, except \jmerger, find very similar mean merger rates for
progenitor ratios larger than $1/3$. In the \hires\ simulation they
also find slightly higher mean merger rates for progenitor ratios
larger than $1/100$. This suggests that in the \hires\ simulation, the
merger rate increases slightly due to the increased number of small
\halos, in agreement with \Fig{pic:Convergence:S:1}, which showed
that the \hires\ simulation has a larger bushiness.
\jmerger\ produces a higher merger rate because it sometimes links the
wrong progenitor to descendant halo, an act which mimics a merger.

\section{Discussion and Conclusions} \label{sec:discussion}

Following our first paper \citep{Srisawat13} and a series of articles
comparing various aspects of merger tree-building codes, we utilized
nine different algorithms to investigate the influence of output
strategy and resolution on the quality of the resulting merger trees.

The output strategy mainly affects the main branch length of the
constructed merger trees. As our results show, somewhat
counterintuitively, increasing the number of outputs from which the
tree is generated results in shorter trees. This is because, due to
limitations in the input halo catalogue, tree-builders may face
difficulties caused by the fluctuating center and size (see for
instance \cite{Srisawat13}, Figure 4) of the input \halos. During
merging events some \halos\ may even disappear completely and then
reappear again in a later snapshot (e.g. \cite{Behroozi2015}, Figure 4). This ambiguous identification will
happen more frequently if there are more snapshots, and this will
increase the chance of terminating a tree main branch prematurely.
This issue is not so prevalent for all our algorithms. It is
particularly bad for class 1 type builders such as \jmerger\ which
lack particle information to aid the halo identification. All four of
our class 2 finders suffer significant problems as the number of
snapshots is increased with halo main branches becoming shorter and
shorter. Of the class 3 finders, where attempts are made to patch over
gaps in the halo history by looking at additional snapshots, the
stability of the reconstruction is varied. Both \sublink\ and
\dtree\ show some residual dependence on snapshot number while
\consistenttree\ is essentially independent of the number of
outputs. \hbt\ is somewhat different, as it is a tracking finder that
interleaves the halo finding and tree-building stages. This generates
a different final halo catalogue which contains more \halos. While the
\MO, \EO\ and \QO\ output strategies display very similar results, the
full \TO\ dataset has a somewhat different dependence due to the
\hbt\ method losing track of small main \halos\ with such finely spaced
outputs.

We also explore the influence of output strategy on bushiness,
a measure of the average number of branches a tree has. We reach the conclusion that this
property changes little with output strategy even though the
corresponding main branch length fluctuates.

Our mass resolution study indicates that, as expected, tree-builders
will build slightly longer trees for \halos\ with the same mass in a
higher resolution simulation. This is true for all our algorithms
except \jmerger\ where the additional \halos\ found at higher
resolution introduce confusion due to the lack of particle ID tracking
which acts to shorten the main branch length. The numerical
resolution of the simulation has a larger influence on the bushiness of the derived
merger trees: higher resolution results in larger bushiness. This
extra bushiness results from the minor progenitors of \halos, as
\Fig{pic:Convergence:P} shows. The number of major mergers does not
change too much. This result is to be expected because, in a low
resolution simulation, very small \halos\ cannot be resolved by the halo
finder. In a higher resolution simulation these small \halos\ appear
and are linked to the branches of the merger tree, resulting in an
increase of the main branch length and the bushiness of the tree. This
resolution dependency mostly comes from the resolution limitation of
the input halo catalogue rather than the tree-builders
themselves. This results in all tree-builders, except \jmerger, producing
very similar results.

As well as investigating the merger tree geometry we also looked into
the mass history and merger rate of our trees. We found, as
for the main branch length, that mass accretion was slightly lower with
many outputs because some trees have been ended prematurely by
occasional dropouts in the halo catalogues.  The mass
history was slightly boosted in the higher mass resolution simulation,
since the finely resolved small \halos\ could extend the trees
branches to higher redshifts.  The merger rate is analogous to
bushiness and was also largely independent of output strategy.

Our results show that patching schemes can improve merger trees. They
also show that complete halo catalogues play an important role in
building merger trees. Such an influence of the input halo catalogue
on merger-tree-building has been discussed by \cite{Avila_2014}. In
this work, we allowed \consistenttree\ and \hbt\ to modified the
initial halo catalogue because it's part of their algorithm. Thus
\consistenttree\ and \hbt\ show the influence of both a patching
scheme and changing the input halo catalogue at the same time.

To conclude, the simulation output strategy chiefly effects the main
branch length of the resultant merger trees. The underlying
simulation's resolution has an effect on both the length and bushiness of the
merger trees. We recommend:

\begin{itemize}
\item Halo merger-tree-builders that do not consider the particle IDs
  should be avoided. They construct trees that do not reflect the
  underlying cosmological model accurately, having an incorrect merger
  rate and typical object age, for example.
\item As has been found previously by \cite{Srisawat13}, all four of
  our class 2 finders, which do not attempt to patch the underlying
  merger tree, are functionally identical. They differ in terms of the
  details of the merit function used to connect \halos\ when building
  the tree. Although this choice makes occasional minor differences
  such that the trees produced are not actually identical, these
  differences are to all intents and purposes irrelevant.
\item Merger trees built from of order 100 or more snapshots should
  always be constructed using an algorithm capable of dealing with
  problems in the underlying halo finder such as missing \halos. This
  patching should ideally be based on a physical timescale rather than
  a fixed number of snapshots and this timescale should be chosen to
  exceed the timescale over which \halos\ typically disappear for.
\item To facilitate this patching at the end of the simulation,
  snapshots should be generated {\it beyond} the desired endpoint. This
would entail typically running past $z=0$. Also, without patching we
have shown in \Fig{pic:missed} that any halo catalogue will be
incomplete at least at the 1 percent level, rising to 2 percent for 64
snaphots. This may be an issue for precision work.
\item Sequences of snapshots with very rapidly changing time intervals
  between them should be avoided as they can lead to very poor trees.
\item \class\ finders, which do not attempt to patch input halo catalogues,
construct merger trees whose main branches are somewhat shorter than those that could
be achieved if patching was applied. As such the merger trees produced
do not accurately reflect the true structure of the underlying
cosmological model. This can be important if the generated merger tree is to be subsequently
used by a semi-analytic model, even though, as
\cite{Lee_2014} found, SAMs can be re-tuned to adjust for an
incomplete tree structure (as this effectively necessitates tuning to
an incorrect cosmology). Well constructed tree-builders capable of
bridging incomplete halo catalogues or ideally fully integrated
tracking finders of class 3 or 4 are therefore preferred.
\end{itemize}

\section*{Acknowledgements} \label{sec:Acknowledgements}
This work was supported by the NSFC projects (Grant Nos. 11473053, 11121062,
11233005, U1331201), the National Key Basic Research Program of China (Grant
No. 2015CB857001), and the ``Strategic Priority Research Program the Emergence
of Cosmological Structures'' of the Chinese Academy of Sciences (Grant No.
XDB09010000).
Part of the simulations in this paper were performed on the High Performance
Computing (HPC) facilities at the University of Nottingham
(www.nottingham.ac.uk/hpc). This work also made use of the High Performance
Computing Resource in the Core Facility for Advanced Research Computing at
Shanghai Astronomical Observatory.

YW is supported by the EC framework 7 research exchange programme LACEGAL.
AK is supported by the {\it Ministerio de Econom\'ia y Competitividad} (MINECO)
in Spain through grant AYA2012-31101 as well as the Consolider-Ingenio 2010
Programme of the {\it Spanish Ministerio de Ciencia e Innovaci\'on} (MICINN)
under grant MultiDark CSD2009-00064. He also acknowledges support from the {\it
Australian Research Council} (ARC) grants DP140100198. He
further thanks Azure Blue for the catcher in the rye.
PAT acknowledges support from the Science and Technology Facilities Council (grant number ST/L000652/1).
PJE is supported by the SSimPL programme and the Sydney Institute for Astronomy
(SIfA) via DP130100117.
SKY acknowledges support from the National Research Foundation of Korea(Doyak
Numerical simulations were performed using the KISTI supercomputer under the programme of KSC-2012-C3-10. 2014003730).

The authors contributed in the following ways to this
paper: YW \& FRP designed the comparison and performed
the analysis. YW is a PhD student supervised by WPL.
YW \& FRP wrote the paper. PAT, CS, AK,
FRP \& AS organized the Sussing Merger Trees workshop at
which this work was initiated. The other authors (as listed in
section \ref{sec:treebuilder}) provided results via their respective tree-building
algorithms. All authors also helped proof-read the paper.

\bibliography{mn-jour,Haloes} \bibliographystyle{mn2e}
\label{sec:Bibliography}
\bsp
\label{lastpage}

\end{document}